\documentclass[]{aastex631}
\usepackage{comment}
\usepackage{xcolor,graphicx,lipsum}
\usepackage[caption=false]{subfig}
\usepackage{ multirow}
\usepackage{float}

\newcommand{\changes}[1]{{\color{red} }}
\newcommand{\astr}{\textit{AstroSat}}
\begin{document}

\title{Spectral evolution and secular shift of the Z source GX 340+0: an interpretation in the spreading layer paradigm}

\author{Aman Kaushik}
\email{aman.kaushik@tifr.res.in}
\affiliation{Department of Astronomy and Astrophysics, Tata Institute of Fundamental Research, 1 Homi Bhabha Road, Colaba, Mumbai 400005, India} 
\author{Yash Bhargava}
\email{yash.bhargava@inaf.it}
\affiliation{INAF - Osservatorio Astronomico di Cagliari, via della Scienza 5, 09047 Selargius (CA), Italy}

\author{Sudip Bhattacharyya}
\email{sudip@tifr.res.in}
\affiliation{Department of Astronomy and Astrophysics, Tata Institute of Fundamental Research, 1 Homi Bhabha Road, Colaba, Mumbai 400005, India}

\begin{abstract}
We present a comprehensive spectral and timing study of the bright neutron star low-mass X-ray binary and Z source, GX~340+0, using broadband X-ray observations with \textit{AstroSat} in 2020 and 2024, focusing on the evolution of physical parameters along the Z track and the origin of secular shifts of this track across different epochs. 
The spectra are consistently described by the same model, comprising blackbody emission from a spreading boundary layer near the neutron star surface, an accretion disk, and a Comptonizing corona. 
We observe a significant decrease in the blackbody temperature from $\sim$1.2 keV on the horizontal branch (HB) to $\sim$1.0 keV on the normal branch (NB), which likely contributes to the harder spectra on the HB and softer spectra on the NB. At the same time, the blackbody normalization varies substantially from $\sim$440--860, indicating changes in the effective emitting area. Within the spreading layer framework, this evolution reflects changes in the extent of the radiating layer, while the large absolute normalization values suggest an additional contribution from the boundary layer.
We also detect variations in the absorption column density, suggesting structural changes in the inner accretion disk, possibly linked to radiation driven puffing-up. The Comptonized flux shows a branch-dependent behavior, remaining relatively stable along the HB but decreasing along the NB. We interpret these results within a unified framework where the boundary layer, corona, and inner disk evolve coherently with accretion rate, explaining both the Z-track behavior and its secular shifts.
\end{abstract}

\section{Introduction}
Neutron star low mass X-ray binaries (NS LMXBs) are broadly classified into two main categories: atoll and Z sources, based on the characteristic patterns traced in their hardness intensity diagram (HID) or color--color diagram (CCD; \citet{1989A&A...225...79H}). Z sources represent the most luminous class of NS LMXBs and typically accrete at rates close to the Eddington limit. In these systems, the spectral evolution produces a characteristic three-branched track in the HID/CCD that resembles the letter ``Z''. The three branches are known as the horizontal branch (HB), the normal branch (NB), and the flaring branch (FB), intersecting at two vertices referred to as the hard apex (HB--NB junction) and the soft apex (NB--FB junction). Z sources also display strong and systematic variations in rapid X-ray variability along the Z track, linking spectral and timing behavior to changes in the accretion flow geometry \citep{2004astro.ph.10551V}. Based on the morphology of the Z track and the relative length of the branches, these sources are further divided into two subclasses: Cyg-like systems (e.g. Cyg X-2, GX 340+0, GX 5-1) and Sco-like systems (e.g. Sco X-1, GX 17+2, GX 349+2; \citet{1994A&A...289..795K, 2010ApJ...719..201H}). Cyg-like sources generally exhibit a prominent HB and relatively short FB, while Sco-like sources display a weaker HB and a more extended FB. In addition to the three canonical branches, some systems occasionally show an additional segment termed the extended flaring branch (EFB), which is thought to be associated with strong instabilities in the inner accretion flow or boundary layer. The HB corresponds to the hardest spectral state of the system, whereas the FB represents the softest state \citep{2013ApJ...766...63S, 2023ApJ...955..102B}. Early studies proposed that motion along the Z track corresponds to a monotonic increase in the mass accretion rate ($\dot{M}$) from HB to FB \citep{1986ApJ...306L..91P}. However, observations showed that the X-ray intensity decreases along the NB despite the expected increase in $\dot{M}$, indicating that the Z-track behavior cannot be explained solely by changes in accretion rate. Consequently, more sophisticated models have been proposed in which the observed spectral evolution is governed by variations in the structure of the inner accretion disk, the NS boundary layer, and the Comptonizing corona \citep{balucinska2010A&A...512A...9B, 2012A&A...546A..35C}. The stability of the Z-track shape over long timescales suggests that these spectral states represent a well-defined sequence of physical conditions within the accretion flow. In order to understand the evolution of the source, its position along the Z track is parameterized using the $S_z$ coordinate, which represents a curvilinear distance measured along the track in HID/CCD \citep{1989A&A...225...79H, 1994A&A...289..795K, 2000ApJ...537..374J, 2000MNRAS.311..201D}. This parameter is constructed by tracing the Z-shaped track and projecting each data point onto the curve, thereby assigning a monotonic coordinate that describes the source evolution along HB, NB, and FB. By convention, $S_z$ increases from the HB through the hard apex toward the NB and FB, providing a convenient proxy for the accretion state of the system.

Long-term monitoring of several Z sources has shown that the characteristic Z-shaped track in HID and CCD is not stationary but can undergo gradual shifts on timescales of days to months while preserving its overall morphology. Such secular shifts were first reported in sources such as GX~5$-$1 and Cyg~X$-$2 using EXOSAT and Ginga observations \citep{1990A&A...235..131H,1994A&A...289..795K}. These studies demonstrated that although the source continues to move along the canonical horizontal, normal, and flaring branches, the entire Z pattern can shift to different intensity and color levels. Subsequent analyses suggested that motion along the Z track reflects short-term changes in the instantaneous mass accretion rate, whereas the secular shift of the track likely results from longer term variations in the average accretion rate or structural changes in the accretion flow \citep{1995A&A...303..801K,2002ApJ...568..878H,2012ApJ...756...34L}.

GX 340+0 is one of the brightest and most extensively studied Cyg-like Z sources in the Galactic plane. The source was identified early on as a Z source through the characteristic Z-shaped pattern observed in its spectral evolution and correlated timing behavior \citep{1988A&A...192..147V, mitsuda1989PASJ...41...97M}. Since then, numerous studies using missions such as \textit{EXOSAT}, \textit{RXTE}, \textit{BeppoSAX}, with more recent and extensive observations from \textit{NICER} and \textit{AstroSat} have investigated its spectral and timing properties in detail. GX 340+0 traces out a full Z track in the HID on timescales of only a few days, covering all three canonical branches and occasionally exhibiting an extended flaring branch \citep{Jonker1998ApJ...499L.191J, Jonker2000ApJ...537..374J, 2024MNRAS.528.4125P}. The source shows a variety of quasi-periodic oscillations (QPOs) associated with different regions of the Z track, including HB oscillations (HBOs) with frequencies of tens of Hz on the HB and NB oscillations (NBOs) around a few Hz along the NB \citep{Jonker2000ApJ...537..374J, Gilfanov2003A&A...410..217G, lavagetto2004NuPhS.132..616L}. Recent observations have also reported HBO frequencies around $20$--$50$ Hz and NBO frequencies of $5$--$8$ Hz, highlighting the strong coupling between spectral state and rapid variability in this system \citep{2023ApJ...955..102B}.

Spectral studies of GX 340+0 generally describe its emission as a combination of thermal and Comptonized components associated with the accretion disk and the boundary layer around the NS \citep{mitsuda1989PASJ...41...97M,Gilfanov2003A&A...410..217G, 2006ChJAS...6a.257I}. Early EXOSAT analyses modeled the spectrum using a blackbody component together with Comptonized emission, with electron temperatures of a few keV \citep{1993A&A...273..123S}. Later broadband studies with BeppoSAX showed that the $0.1$--$200$ keV spectrum can be well described by a soft blackbody component with $kT \sim 0.5$ keV together with a Comptonized component with electron temperatures of a few keV, and an additional power-law tail at higher energies \citep{2006ChJAS...6a.257I}. An Fe~K$\alpha$ emission line around $6.7$--$6.9$ keV has also been detected in several observations, indicating reflection from the inner accretion disk \citep{lavagetto2004NuPhS.132..616L, dai2009ApJ...693L...1D}. Alternative spectral interpretations have been proposed as well. In the extended accretion-disk corona model, the emission consists of blackbody radiation from the NS surface together with Comptonized emission produced in an extended corona surrounding the disk \citep{balucinska2010A&A...512A...9B, 2012A&A...546A..35C}. On the other hand, \citet{Seifina2013ApJ...766...63S} proposed a model involving multiple Comptonization regions, including a transition layer between the accretion disk and the NS surface. These differing interpretations highlight the ongoing debate regarding the dominant emission components and their evolution along the Z track. In addition to its X-ray spectral and timing properties, several aspects of the binary geometry and accretion environment of GX 340+0 remain uncertain. Recent X-ray polarimetric observations from \textit{IXPE} have begun to provide some constraints on the geometry of the inner accretion flow, indicating that the polarized emission likely originates from a Comptonizing region, preferably of slab-like geometry rather than the optically thick thermal components. The relatively low polarization degree and nearly constant polarization angle across different branches suggest a stable global geometry, while the variation in polarization strength points to changes in the relative contribution of the Comptonizing medium and its coupling with the neutron star surface and inner disk \citep{2024arXiv240519324B, 2024arXiv241100350B, 2024A&A...691A.253L, 2025A&A...702A.101L, 2025A&A...699A.230G}.
The binary inclination of the system is not well constrained due to the absence of direct optical measurements, primarily because the source lies in a highly absorbed region of the Galactic plane. Nevertheless, indirect constraints on the inclination have been obtained from X-ray observations. The absence of X-ray dipping behavior, which is commonly observed in high inclination systems where the accretion disk periodically obscures the central source, suggests that the inclination angle is likely below $\sim 70^\circ$ \citep{frank87}. More quantitative estimates have been obtained through reflection modeling of the Fe~K$\alpha$ emission line profile originating from the inner accretion disk. Such analyses indicate a moderately inclined geometry with $i \approx 37^\circ$ \citep{dai2009ApJ...693L...1D, miller2016ApJ...822L..18M, monaca2024A&A...691A.253L}. These reflection studies also provide important constraints on the structure of the inner disk and the relativistic effects shaping the observed line profile. In addition, previous studies have shown secular shift in the HID for GX~340+0. These long-term variations have also been reported earlier, where different observations show similar Z track morphology but at different overall intensity levels, indicating changes in the long-term accretion environment of the system \citep{1998ApJ...499L.191J, 2002A&A...386..535D}. 

In this work, we present a detailed spectral and aperiodic timing study of the NS Z source GX~340+0 using \textit{AstroSat} observations obtained during 2020 and 2024. The primary aim of this study is to investigate the physical mechanisms governing the evolution of the source along the characteristic Z track in the HID. In particular, we focus on understanding the factors that cause the source to remain confined to the Z-shaped track by examining the systematic variation of spectral parameters along different regions of the track. To quantify this evolution, the Z track is parameterized using the $S_z$ coordinate and divided into several segments, allowing us to perform a consistent spectral analysis across the entire track. This systematic analysis using $S_z$ values has been done for the first time for this source using \textit{AstroSat} data. The same spectral model is applied to all segments in order to investigate how the properties of the emission components evolve as the source moves along the Z track. In addition, we explore the long-term secular shifts observed in the HID by comparing observations obtained over different epochs. Such shifts may provide important insights into possible changes in the accretion flow geometry or global accretion rate over time. 
The structure of the paper is as follows. In Section~\ref{sec:obs_analysis} we describe the \textit{AstroSat} observations and the data reduction procedures. The spectral analysis and corresponding results are presented in Section~\ref{sec:spectral_results}. In Section~\ref{discussion} we interpret our findings and discuss their implications for the accretion processes in GX~340+0. Finally, we summarize the main conclusions of this work in Section~\ref{sec:summary}.

\section{Observation and Data analysis} \label{sec:obs_analysis}

\subsection{AstroSat/LAXPC}

\textit{AstroSat}, launched in 2015, is India's first dedicated multi-wavelength astronomy satellite \citep{2006AdSpR..38.2989A, 2014SPIE.9144E..1SS}. It is equipped with four main scientific pointing payloads including the Large Area X-ray Proportional Counters \citep[LAXPC; ][]{2016SPIE.9905E..1DY,Yadav2016SPIE.9905E..1DY,Yadav2017CSci..113..591Y,2017ApJS..231...10A} and the Soft X-ray Telescope \citep[SXT; ][]{2016SPIE.9905E..1ES, 2017JApA...38...29S}.

\textit{AstroSat}/LAXPC has three identical co-aligned proportional counters operating in the 3–80 keV energy range named LAXPC10, LAXPC20, and LAXPC30. Each \textit{AstroSat}/LAXPC detector consists of five layers and independently records the arrival time of each photon with a time resolution of 10\,$\mu$s and dead time of 42\,$\mathrm{\mu}$s. The orbit-wise level 1 data is downloaded from AstroBrowse\footnote{\url{https://astrobrowse.issdc.gov.in/astro_archive/archive/Home.jsp}} with observation ID: T05$\_$188T01$\_$9000006138 and T05$\_$218T01$\_$9000006386. 
For our analysis, we use only the LAXPC20 detector as LAXPC10 has been operating at low gain since 2018, and detector LAXPC30\footnote{\url{http://astrosat-ssc.iucaa.in/}} had been switched off early in the mission due to a leak in the detector. 
To minimize the background, only the top layer (L1) data are used, which restricts the operational energy range to 3--20~keV.
We reduce the \textit{AstroSat}/LAXPC data using the pipeline \texttt{LAXPCSOFTWARE22AUG15}\footnote{\url{http://astrosat-ssc.iucaa.in/uploads/laxpc/LAXPCsoftware22Aug15.zip}} \citep{Antia2021JApA...42...32A, Misra2021JApA...42...55M}. The software includes
tools to reduce the level 1 data to level 2 data, calibration files, responses, etc. We obtain the level 2 data and standard good time intervals (GTIs) using the tools from the pipeline. 

\subsection{AstroSat/SXT}

The orbit-wise level 2 processed data was procured from AstroBrowse\footnote{\url{https://astrobrowse.issdc.gov.in/astro_archive/archive/Home.jsp}}  which has been processed with the latest version of the SXT\footnote{\url{https://www.tifr.res.in/astrosat_sxt/index.html}} pipeline (AS1SXTLevel2-1.5). 
Individual orbits were merged using the \textsc{Julia} tool \texttt{SXTMerger} as suggested by the SXT Payload Operation Center (POC). The standard products (i.e. spectra and light curves) were extracted using \texttt{xselect} (v2.5b from \textsc{heasoft} v6.32.1) for a circular region of 14$\arcmin$. The last three orbits of the SXT March observations showed abnormal excess in count rate, which is not seen in the same bands in LAXPC light curves. Thus, we excluded those orbits from both LAXPC and SXT spectra \citep{2024arXiv240519324B}. Table~\ref{tab:obBLog} shows the exposure time excluding those orbits. For the extraction of the spectral products, only the grade 0 events were selected, and the appropriate response matrix was obtained from the POC. The standard ancillary response file was modified for the selected region using the \texttt{sxtARFModule} tool provided by the POC. We used \texttt{ftgrouppha} to bin the spectra using response files for optimum binning.

\begin{table*}
  \centering
  \caption{\textit{AstroSat} observation log table for 
the source GX 340+0 with net exposure times.}\label{tab:obBLog}
  \begin{tabular}{|ll|c|ccc|}
    \hline
    Observatory & Instrument & Observation ID & Start Time & Stop Time & Exposure (in ks) \\ \hline
    \multirow{2}{*}{\astr} & SXT & \multirow{2}{*}{9000006138} & 2024-03-23 12:23:48 & 2024-03-24 00:15:28 & 14.2 \\
     & LAXPC & & 2024-03-23 12:24:50 & 2024-03-24 00:13:10 & 23.1  \\ \hline
    \multirow{2}{*}{\astr} & SXT & \multirow{2}{*}{9000006386} &2024-08-13 12:48:03 & 2024-08-14 06:47:02 & 13.3 \\
    & LAXPC &  & 2024-08-13 13:07:31 & 2024-08-14 07:01:41& 16.4 \\ \hline 
\end{tabular}
\end{table*}

\section{Spectral analysis and results} \label{sec:spectral_results}

\subsection{Hardness Intensity Diagram} \label{subsec:hid}
\begin{figure*}
    \centering
\includegraphics[width=0.49\columnwidth]{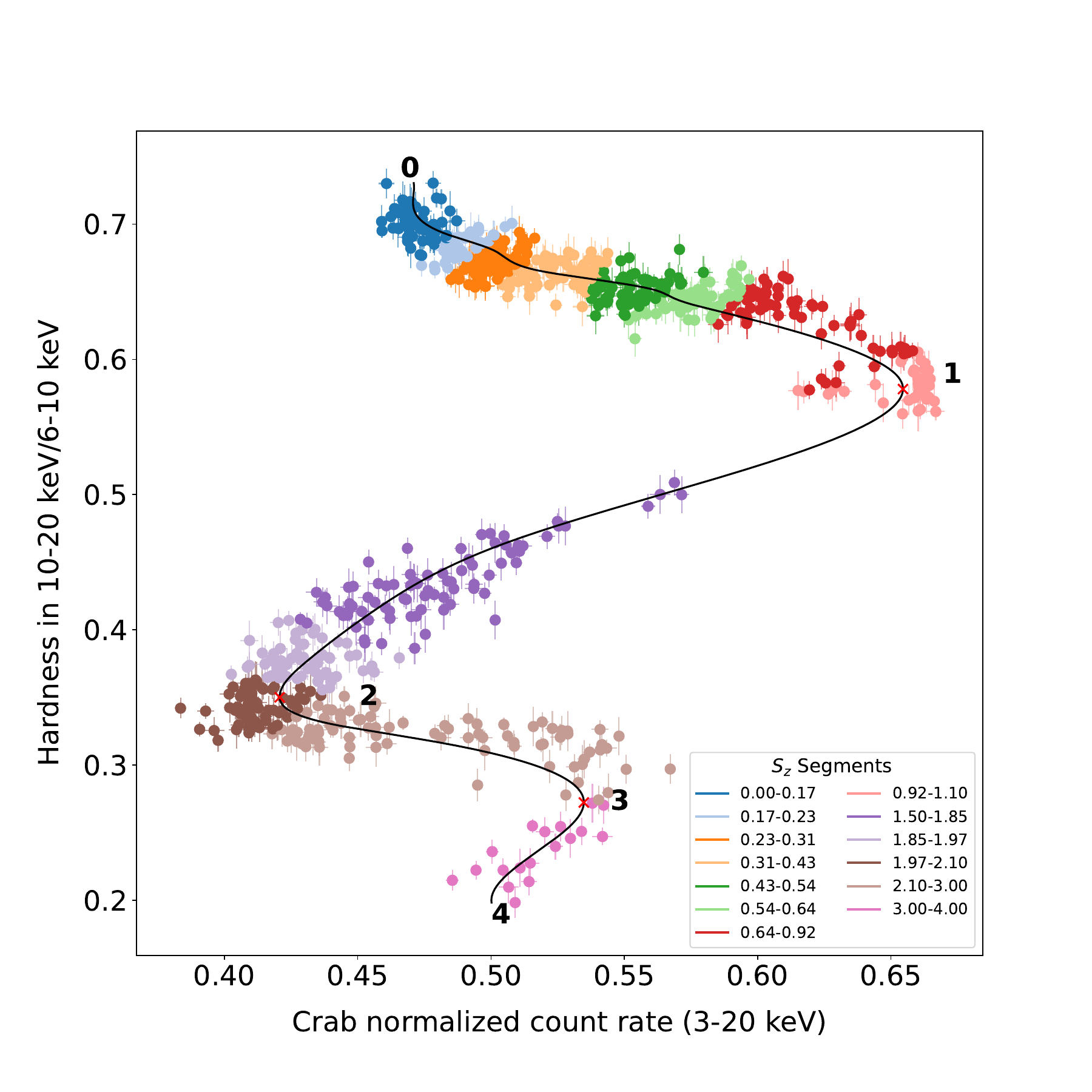}
\includegraphics[width=0.49\columnwidth]{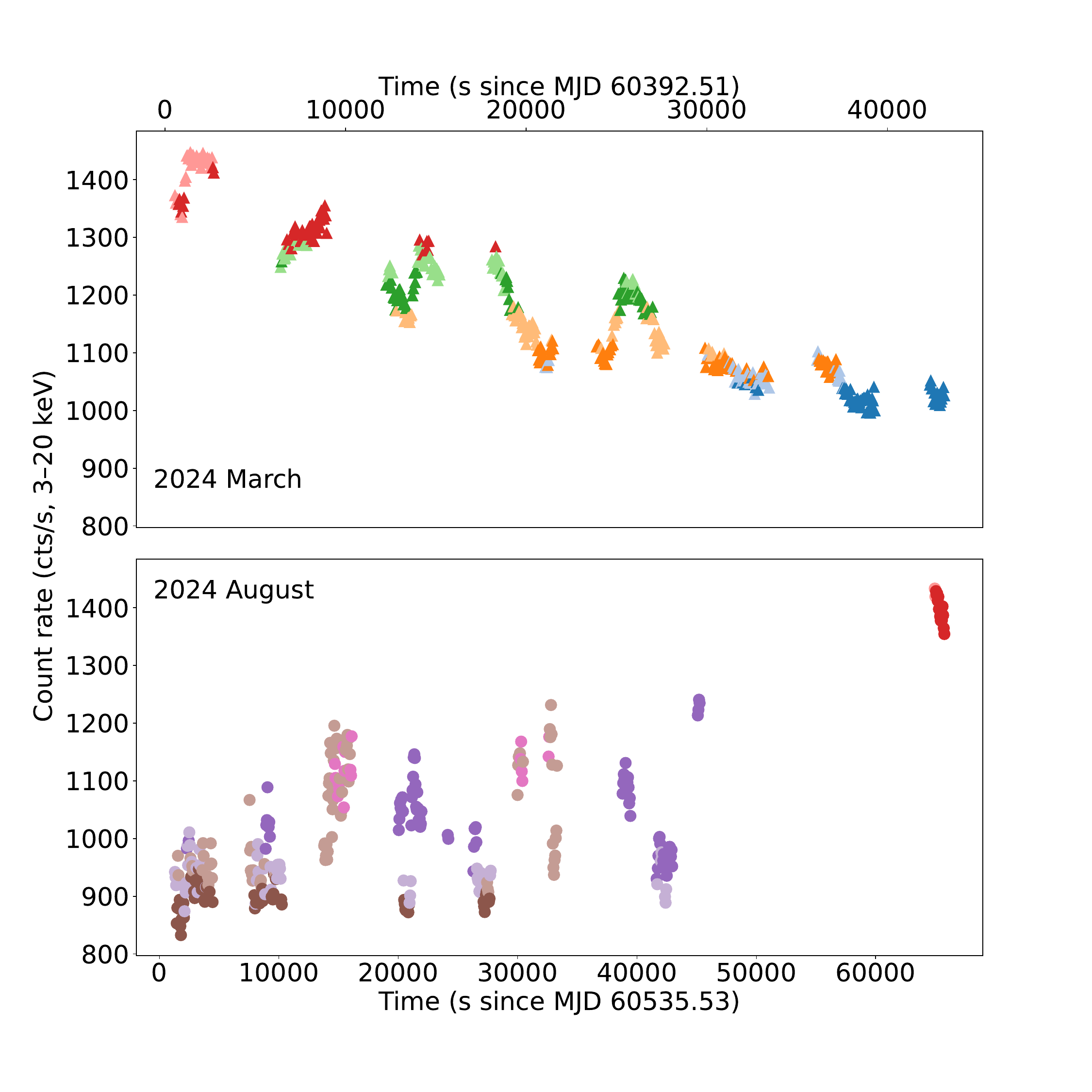}
\caption{The Figure shows the hardness intensity diagram (HID) of source GX 340+0 observed in 2024 by \textit{AstroSat} (left panel) and lightcurves (right panels). The HID shows that the source has traversed through all the branches during the March and August observations, section~\ref{subsec:hid}. The black line shows the fitted spline through the data. The red color cross marks the vertices of the spline curve with numbers representing the $S_z$ values at the end points and the vertices. We have used the same color coding in the HID and lightcurve for different $S_z$ segments.}

    \label{fig:lightcurve_hid}
\end{figure*}

We extract the 50\,s binned light curve for both the March and August 2024 observation in the 3.0--20.0\,keV energy range for \textit{AstroSat}/LAXPC and depict the evolution of the source in Figure~\ref{fig:lightcurve_hid} right panel. We plot the hardness intensity diagram (HID) by taking the ratio of count rate in 6--10\,keV and 10--20\,keV energy range. The HID indicates that the source traversed through all the branches of the Z-curve in the 2024 observation, see~\ref{fig:lightcurve_hid} left panel. Comparing the HID and light curve, we can infer that the source went into the HB only once during the August observation and was in the NB with a
few excursions into the FB and EFB during the rest of the time (see also \citealt{2024arXiv240519324B, 2024arXiv241100350B}). 

\subsection{Segmenting based on $S_z$ values}

For tracing the evolution of the spectrum along the Z-track, we segmented the Z-curve of GX 340+0 using the March and August 2024 data. The segments are defined based on the $S_z$ values which are assigned to each data point of the Z-curve as follows: we fit the Z-curve with a cubic spline with 15 internal knots to trace out the density variation in the curve. We label the start, end, and bending points of the Z-curve by numbers, e.g., number 0 for the starting point of the HB branch, 1 for the HB-NB apex, 2 for the NB-FB apex, and so on \citep{1994A&A...289..795K, 2000ApJ...537..374J}. Then, we drop a perpendicular line from each data point intersecting the fitted spline. Based on the position of intersection, we associate a $S_z$ value to each data point such that points on HB lie in the range 0--1, points on NB lie between 1--2, and so on. These unique values for the data points are then used to segment the data in each branch of the Z-curve. For our analysis, we use $\approx$60 data points ($\approx$3\,ks exposure) in each segment and fix the segment boundaries for the apexes to avoid mixing of different branches (table~\ref{tab:spec_pars}). Additionally, the upper NB is sparsely sampled for 2024 data, with only six data points corresponding to a total exposure of approximately 300\,s. Therefore, these points are excluded from the analysis, and the ($S_z$) interval is taken from 1.5--1.85, table~\ref{tab:spec_pars}. For the 2020 data, the \(S_z\) values are assigned by first overlapping the 2020 HID with the 2024 HID, taking the hard vertex as the common reference point. The extent of the 2020 track is then identified on the 2024 \(S_z\) scale. From this comparison, we find that the 2020 HID spans the portion of the 2024 track corresponding to \(S_z \sim 0.17\) to \(1.4\). Therefore, the 2020 data is assigned \(S_z\) values over this range. We extracted GTI based on these $S_z$ segment boundaries for generating lightcurves and HID. Since both, LAXPC and SXT are co-pointed with SXT observations more constrained than LAXPC, same GTI is also applied to the SXT observations.

\subsection{Spectral analysis}
\begin{figure*}
    \centering
\includegraphics[width=0.49\columnwidth]{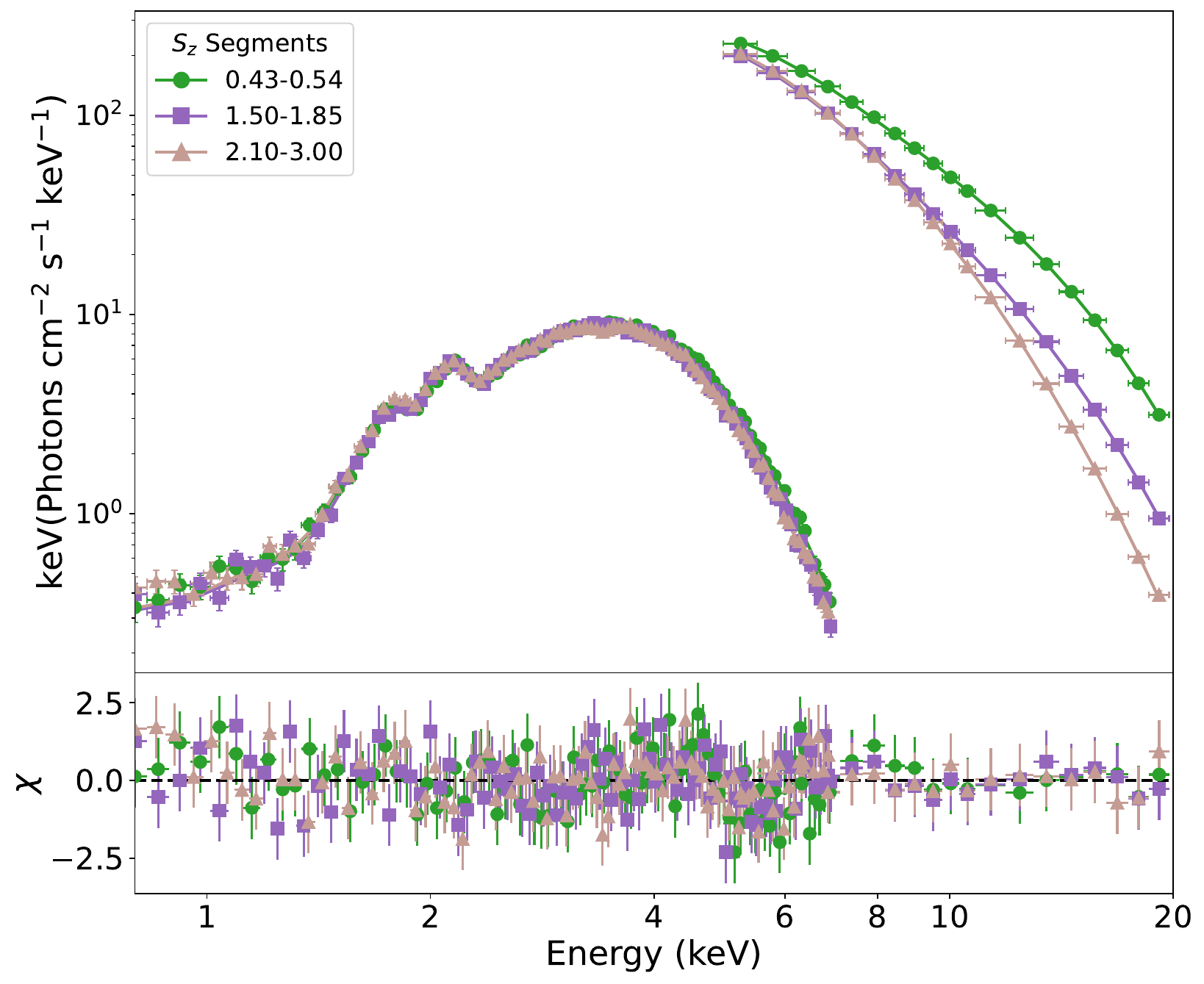}
\includegraphics[width=0.49\columnwidth]{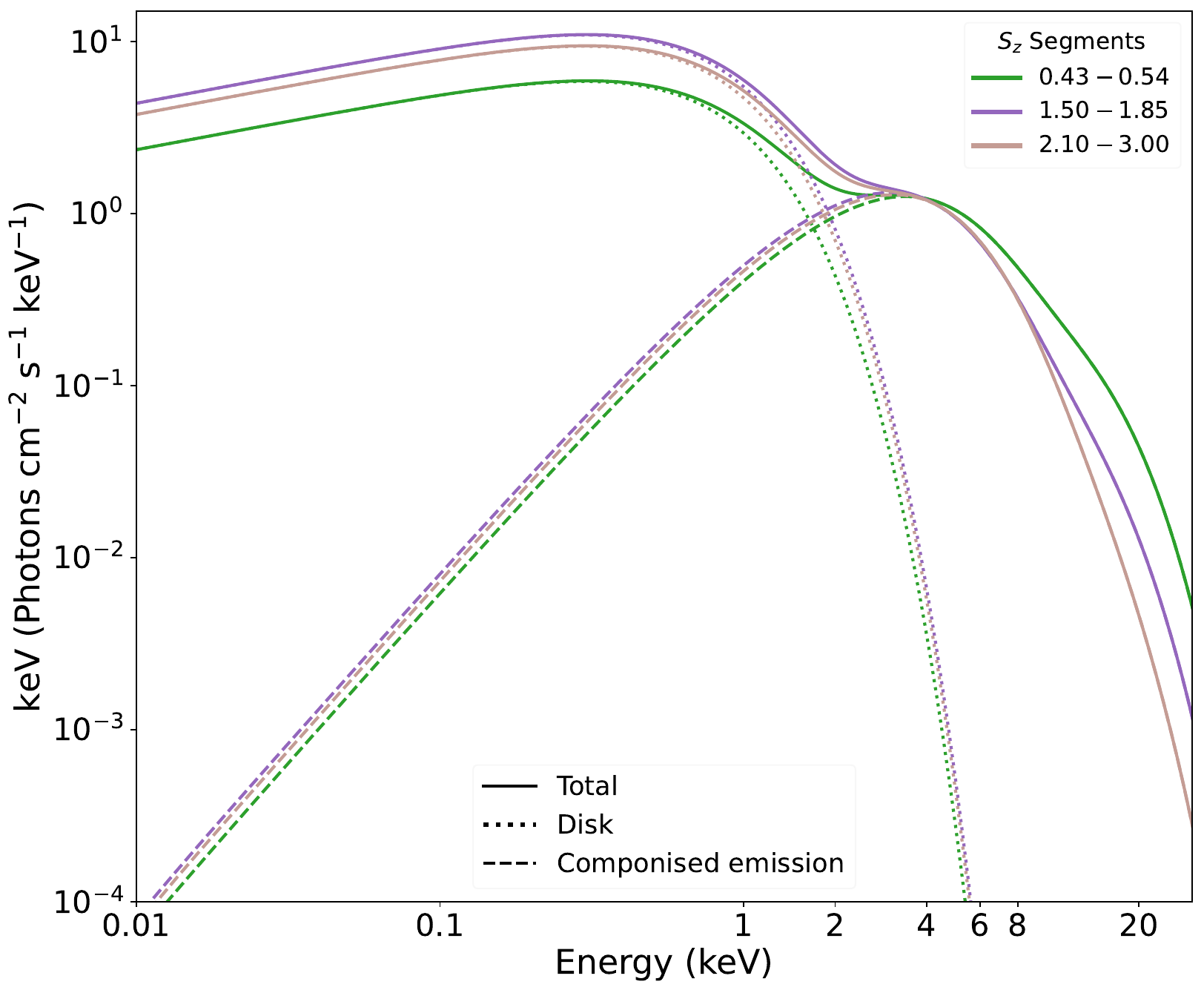}
\caption{Left: Representative spectra in 0.7--20.0\,keV range of GX~340+0 corresponding to three segments of the Z track, illustrating the horizontal branch ($S_z: 0.43-0.54$), normal branch ($S_z: 1.50-1.85$), and flaring branch ($S_z: 2.10-3.00$), as observed with \textit{AstroSat}/SXT and LAXPC in 2024. The lower panel shows the residuals with respect to the best-fit model, \texttt{tbabs*(diskbb+thcomp$\otimes$bbodyrad)}. Right: Unabsorbed spectral decomposition of the best-fit model for the representative HB, NB, and FB spectra. The solid curves represent the total model, while the dotted and dashed curves show the contributions from the disk and Comptonized components, respectively. The color coding is identical to that used in the $S_z$-resolved HID (Figure~\ref{fig:lightcurve_hid})}

    \label{fig:spectrum}
\end{figure*}

We perform spectral fitting for all $S_z$ segments using the same spectral model in order to consistently track the evolution of spectral parameters along the Z track. The best-fit spectral values are given in table~\ref{tab:spec_pars}. Since the exposure time for the extended flaring branch (EFB) is low, $\sim$900\,s (red data points in the HID, Figure~\ref{fig:lightcurve_hid}), the spectral parameters are not well constrained. Hence, we do not depict the EFB best-fit parameters value in the evolution plots. The spectra are fitted using \texttt{XSPEC v12.14.0h}, employing $\chi^2$ statistics, and all parameter uncertainties are quoted at the $1\sigma$ confidence level. To account for absorption by interstellar and intrinsic neutral hydrogen, we used the \texttt{tbabs} model. The continuum emission in all segments is well described by a combination of thermal disk emission (\texttt{diskbb}), blackbody emission from the neutron star surface (\texttt{bbodyrad}), and a Comptonized component (\texttt{thcomp}), where the seed photons for Comptonization are assumed to originate from the neutron star surface. As a representative example, we briefly describe below the spectral fitting of one segment corresponding to $S_z = 0.43$–0.54, while the same fitting prescription is applied to all other segments (see Figure~\ref{fig:spectrum}). 
We consider abundances from \citet{2000ApJ...542..914W} and the cross-sections from \citet{1996ApJ...465..487V} to model the spectrum. 
We limit the spectrum for each segment to an energy range where the source counts are above the background counts (e.g., for the segment with $S_z$ range 0.43--0.54, the considered energy range is 0.7--7.0\,keV for SXT and 4.5--20.0\,keV for LAXPC). We start with the model \texttt{constant*tbabs*(diskbb)} giving an unacceptable $\chi^2$/degrees of freedom(dof) = 191123/104. This high residual in the soft X-ray band is significantly improved by the addition of single temperature blackbody component (\texttt{bbodyrad}) giving $\chi^2$/dof = 270/102. To model the hard tail, we use the \texttt{XSPEC} component \texttt{thcomp} (\texttt{constant*tbabs*(diskbb+thcomp$\otimes$bbodyrad}). The convolution model \texttt{thcomp} is an updated version of \texttt{nthcomp} and agrees with the actual Monte Carlo spectra from Comptonization much better than \texttt{nthcomp} \citep{2020MNRAS.492.5234Z}. 
The component \texttt{thcomp} involves Compton up-scattering fraction (\textit{f}) of photons from any seed emission, and gives the low-energy cut-off corresponding to the soft tail of the seed photon spectrum and the high-energy rollover corresponding to a thermal distribution of electrons with a temperature of k$T_\mathrm{e}$.

Thus, the seed photon source parameters can be simultaneously constrained in such a description (while \texttt{nthcomp} allows for constraint on the seed photon temperature for only two cases, a single temperature blackbody or a multicolor blackbody). For an accurate computation of the convolution model, we use the energy range from 0.01--100\,keV having 1000 logarithmic bins which is significantly larger than the range covered by the instrument response. In order to test if the seed photons could have been supplied by the disk, we tried the model \texttt{constant*tbabs*(bbodyrad+thcomp$\otimes$diskbb)} for all the $S_z$ segments, but the fit was poor in some of the segments. Based on these results, we favor the interpretation in which the seed photons primarily originate from the neutron star surface rather than the disk. Additionally, we do not detect any statistically significant residuals around the Fe K$\alpha$ emission region near $\sim$6.4 keV in our spectra because of LAXPC's low resolution.

\begin{figure*}
    \centering
\includegraphics[width=0.49\columnwidth]{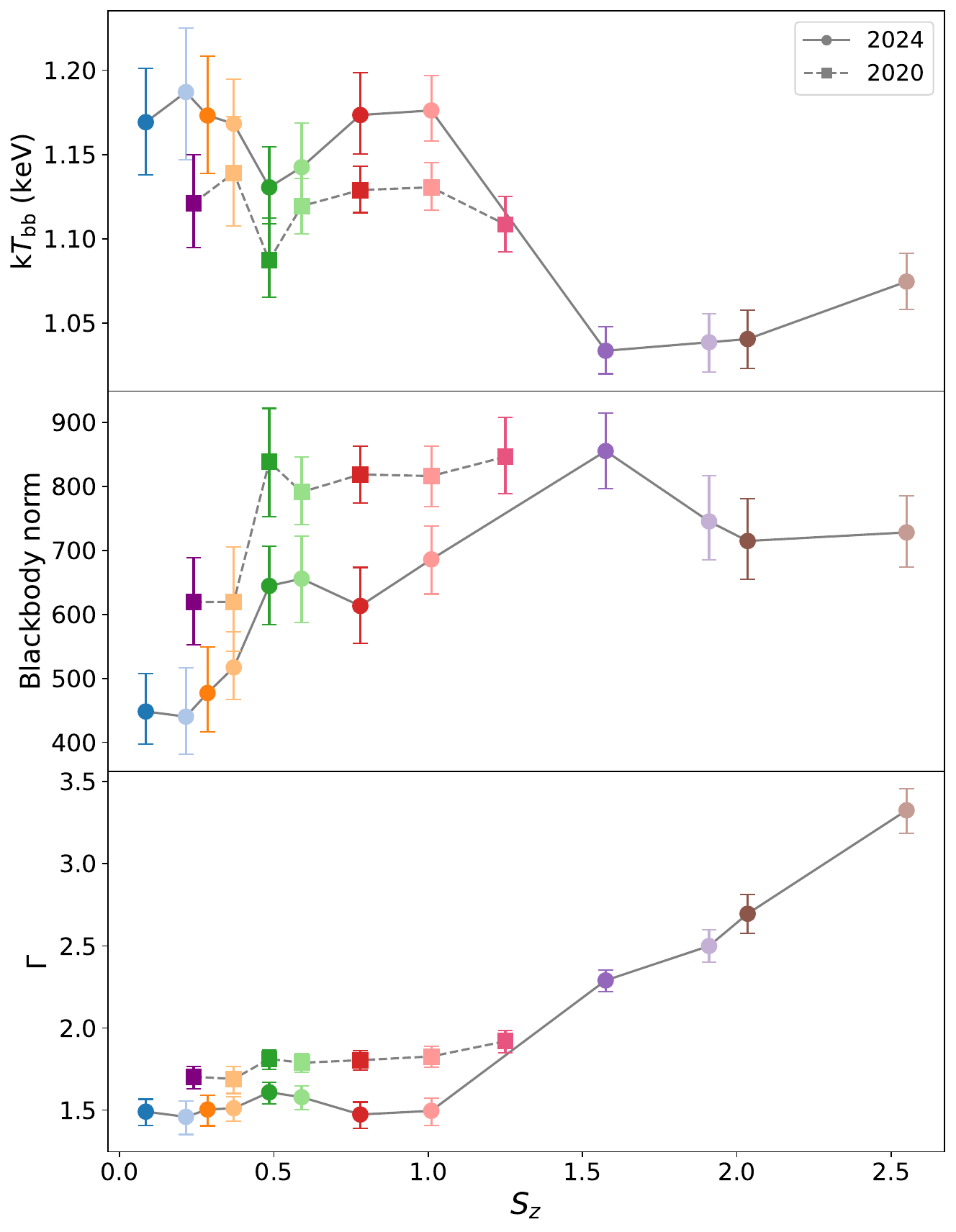}
\includegraphics[width=0.495\columnwidth]{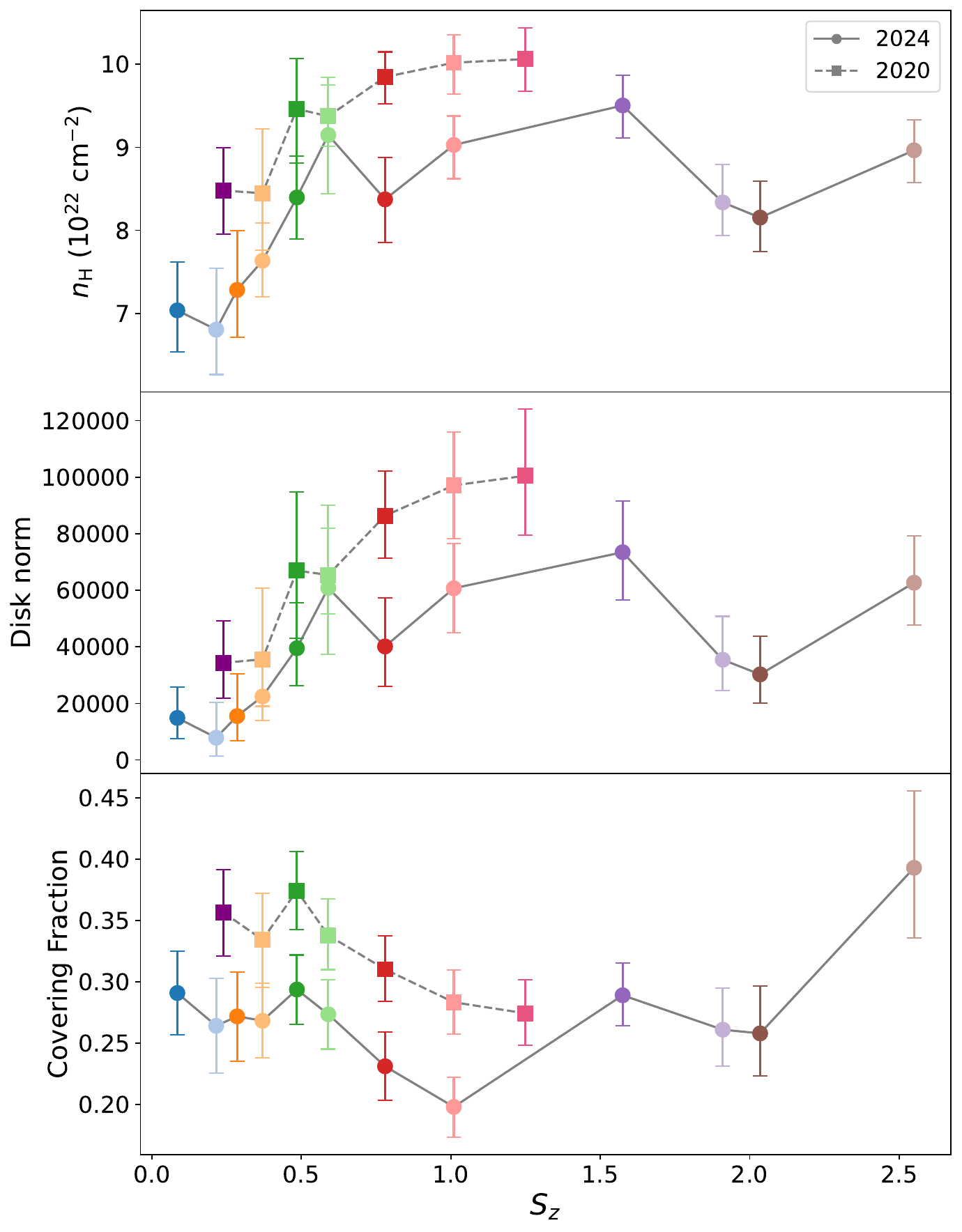}
\caption{Evolution of best-fit spectral parameter values throughout the 2020 and 2024 observation of GX~340+0 by \textit{AstroSat}. The \texttt{XSPEC} model \texttt{tbabs*(diskbb+thcomp$\otimes$bbodyrad}) is applied to all the $S_z$ segments and the data points have the same color coding as the lightcurve and HID (see Figure~\ref{fig:lightcurve_hid}). The first (purple) and last (pink) data points for 2020 data in each panel correspond to $S_z$ range 0.17--0.31 and 1.10--1.40 respectively.}

    \label{fig:parameter_evo}
\end{figure*}

\begin{figure*}
    \centering
\includegraphics[width=0.49\columnwidth]{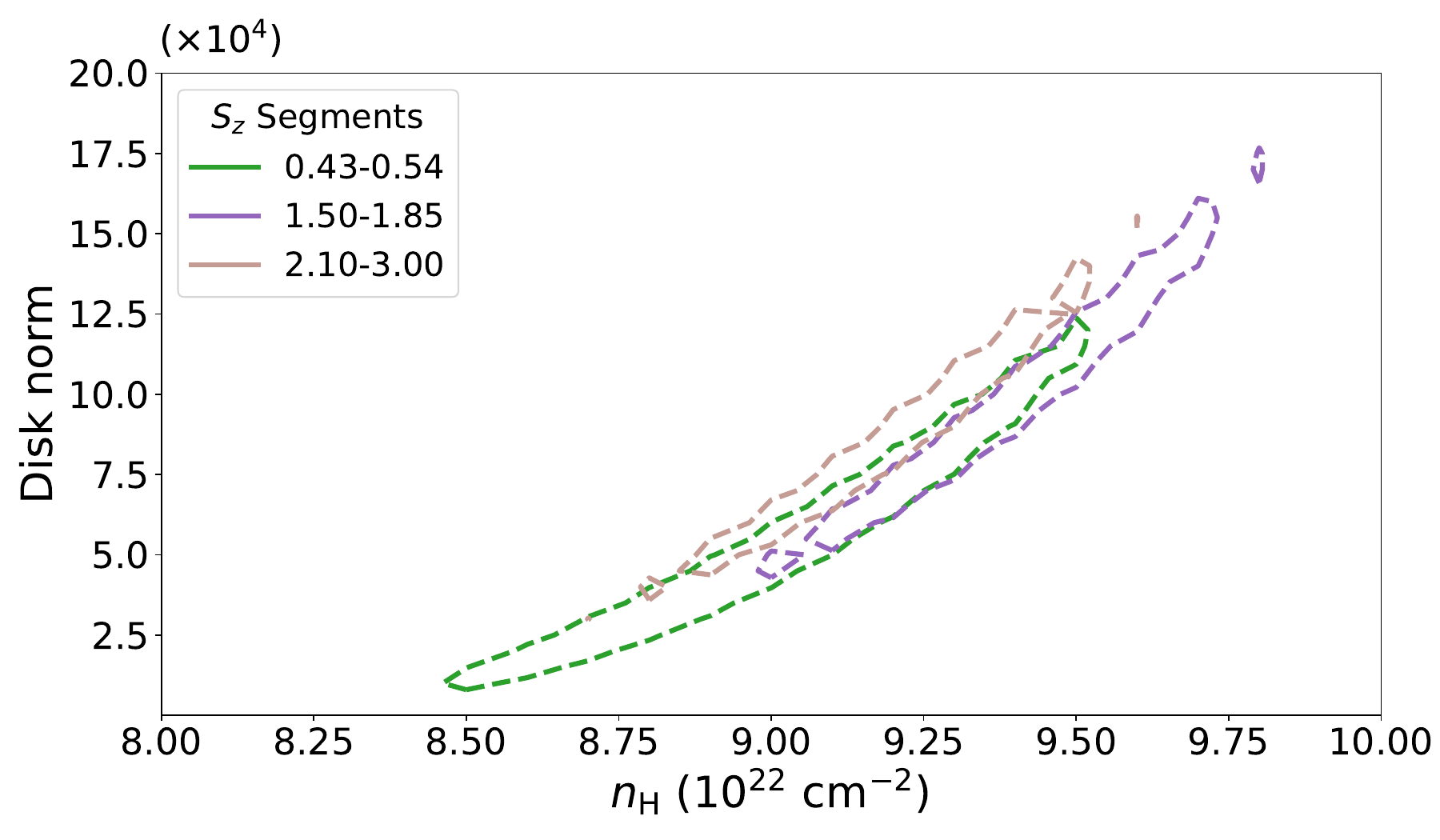}
\includegraphics[width=0.49\columnwidth]{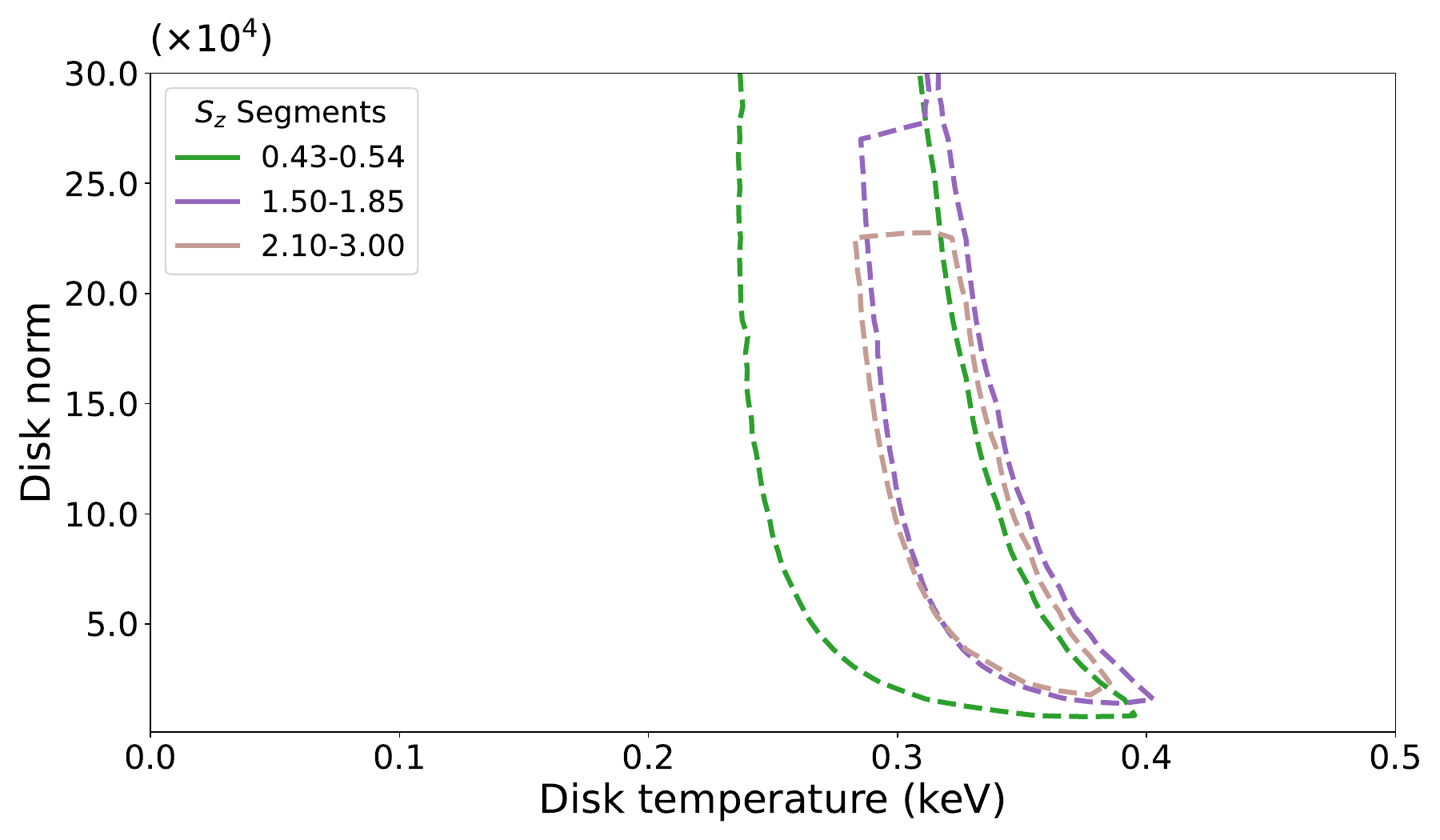}
\caption{$\chi^{2}$ contour curves showing the confidence regions in the Disk norm–$n_{\rm H}$ parameter space (left panel) and Disk norm-Disk temperature space (right panel) for different $S_z$ segments. The contours correspond to the 3$\sigma$ confidence level and demonstrate the degeneracy between these spectral parameters. The color coding is identical to that used in the $S_z$-resolved HID (Figure~\ref{fig:lightcurve_hid}). Although a similar parameter degeneracy is present for all $S_z$ segments, only a representative subset is displayed here to prevent overcrowding of the figure. The lower limit of Disk norm was fixed at 50, as smaller values would imply an apparent inner disk radius smaller than the NS radius ($<$10\,km),  assuming an inclination of 37$^\circ$ and a distance of 11\,kpc \citep{2000MNRAS.317....1F} and is therefore physically implausible.}
    \label{fig:contour_plot}    
\end{figure*}

The disk temperature and disk normalization exhibit a strong covariance, making it difficult to constrain both parameters simultaneously (Figure~\ref{fig:contour_plot}). To minimize this degeneracy, we simultaneously fitted representative spectra from the HB, NB, and FB while tying the disk temperature across all three branches. This yielded a common best-fit value of $\mathrm{k}T_{\rm in}\simeq0.33\pm0.02$ keV, indicating that the characteristic disk temperature remains nearly constant across the Z track. This is further supported by the disk flux--temperature relation, which shows that the disk flux varies while the disk temperature remains confined to a narrow range (see Figure~\ref{fig:appendix_fig}). Such behaviour is unlikely to arise from repeated heating and cooling of the accretion disk on the relatively short timescales over which the source traverses the Z track. Instead, the observed flux variations are more naturally explained by changes in the apparent emitting area associated with the varying disk normalization, consistent with the disk flaring or vertical structural changes discussed in Section~\ref{subsec:spectral_evolution}. Since the peak of such a cool disk lies largely below the observed energy range and is significantly affected by interstellar absorption, the spectra primarily sample only the high energy tail of the disk component, making the disk normalization poorly constrained when both parameters are allowed to vary simultaneously. We therefore fixed $\mathrm{k}T_{\rm in}=0.33$ keV, corresponding to the representative value obtained from the simultaneous fit, to obtain stable and physically self-consistent spectral parameters across all $S_z$ segments. We also repeated the covariance analysis using {\tt diskpn} model \citep{1999MNRAS.309..496G} and observed the similar degeneracy between these parameters.
The electron temperature $\mathrm{k}T_{\rm e}$ was reliably constrained only in a subset of the spectra and showed an average value of $\sim$3.1 keV. In several other $S_z$ segments, however, $\mathrm{k}T_{\rm e}$ became unconstrained, and consequently the photon index $\Gamma$ could no longer be determined reliably. This resulted in unstable fits with large parameter uncertainties. To obtain robust measurements of the spectral evolution along the Z track, we therefore fixed $\mathrm{k}T_{\rm e}$ at the average value of 3.1 keV derived from the constrained spectra. Figure~\ref{fig:parameter_evo} shows the evolution of the rest of the fitted parameters across all the segments with the same color coding as HID. Additionally, to investigate the apparent correlation between $n_\mathrm{H}$ and disk normalization, we generate $\chi^2$ contour plots, which reveal similar test-statistic for many combinations of the parameters (Figure~\ref{fig:contour_plot}). The correlated evolution of both quantities along the Z track suggests that their variations are not entirely independent. This behavior can be interpreted in terms of structural changes in the inner disk, where an increase in vertical scale height (disk puffing) enhances both the line-of-sight absorption and the apparent disk normalization. Therefore, the disk normalization may not solely trace the inner disk radius, but also encode information about the geometrical thickening of the disk.

\begin{table*}
    \centering
    \caption{The best-fit spectral parameter values for the Z source GX~340+0 during its 2020 and 2024 observations with \textit{AstroSat}.
    }\label{tab:spec_pars}
    
    \begin{tabular}{|c|c|c|c|c c|c|c c|c|c|}
    
        \hline
        \multicolumn{4}{|c|}{}& \multicolumn{2}{c|}{thcomp} & \multicolumn{1}{c|}{diskbb} & \multicolumn{2}{c|}{bbodyrad} & & \\ \hline
        
        Obs. & $S_{z}$range$^{\S}$ & Exposure$^{\ast \ast}$ & $n_\mathrm{H}^{\ddag}$ & $\Gamma^{\P}$ & \textit{f}$^{\bullet}$\ & norm$^{\ast}$ & $\mathrm{k}T_{\mathrm{bb}}^{\star}$ & $R_{km}^{\dagger}$ & Flux$^{\#}$ & $\chi^2/$dof \\
        \hline
        \hline
        \multirow{7}{*}{2020} & 0.17--0.31 & 3.6 & 8.5$\pm0.5$ & 1.70$\pm0.06$ & 0.36$\pm0.03$ & 3.4$^{+1.5}_{-1.2}$ & 1.12$\pm0.03$ & 27.4$\pm1.5$ & 10.13$\pm0.08$ & 98/95 \\
        & 0.31--0.43 & 4.4 & 8.4$\pm0.7$ & 1.69$\pm0.08$ & 0.33$\pm0.04$ & 3.6$^{+2.5}_{-1.7}$ & 1.14$\pm0.03$ & 27.4$\pm1.8$ & 10.68$\pm0.08$ & 76/97 \\
        & 0.43--0.54 & 8.3 & 9.5$\pm0.6$ & 1.81$\pm0.06$ & 0.37$\pm0.03$ & 6.7$^{+2.7}_{-2.4}$ & 1.09$\pm0.02$ & 31.9$\pm1.5$ & 11.82$\pm0.09$ & 88/99 \\
        & 0.54--0.64 & 18.5 & 9.4$\pm0.4$ & 1.78$\pm0.06$ & 0.34$\pm0.03$ & 6.5$^{+1.6}_{-1.4}$ & 1.12$\pm0.02$ & 30.9$\pm1.0$ & 12.36$\pm0.08$ & 109/107 \\
        & 0.64--0.92 & 32.8 & 9.8$\pm0.3$ & 1.80$\pm0.06$ & 0.31$\pm0.03$ & 8.6$^{+1.5}_{-1.5}$ & 1.13$\pm0.01$ & 31.5$\pm0.8$ & 12.92$\pm0.08$ & 103/109 \\
        & 0.92--1.10 & 27.5 & 10.0$\pm0.3$ & 1.83$\pm0.06$ & 0.28$\pm0.03$ & 9.7$^{+1.9}_{-1.9}$ & 1.13$\pm0.01$ & 31.5$\pm1.0$ & 12.62$\pm0.09$ & 100/105 \\
        & 1.10--1.40 & 11.5 & 10.1$\pm0.4$ & 1.91$\pm0.07$ & 0.27$\pm0.03$ & 10.0$^{+2.3}_{-2.1}$ & 1.11$\pm0.02$ & 32.1$\pm1.1$ & 11.70$\pm0.09$ & 91/100 \\
        \hline
        \multirow{12}{*}{2024} & 0.0--0.17 & 3.0 & 7.0$\pm0.5$ & 1.50$\pm0.08$ & 0.29$\pm0.03$ & 1.5$^{+1.1}_{-0.7}$ & 1.17$\pm0.03$ & 23.3$\pm1.3$ & 8.81$\pm0.08$ & 99/95 \\
        &0.17--0.23 & 3.2 & 6.8$^{+0.5}_{-0.7}$ & 1.46$\pm0.10$ & 0.26$\pm0.04$ & 0.8$^{+1.2}_{-0.6}$ & 1.19$\pm0.04$ & 23.1$^{+1.8}_{-1.6}$ & 9.07$\pm0.08$ & 89/97\\
        &0.23--0.31 & 3.0 & 7.3$^{+0.6}_{-0.7}$ & 1.50$\pm0.09$ & 0.27$\pm0.04$ & 1.6$^{+1.5}_{-0.9}$ & 1.17$\pm0.03$ & 24.1$^{+1.7}_{-1.5}$ & 9.30$\pm0.07$ & 90/96\\
        &0.31--0.43 & 3.2 & 7.6$\pm0.4$ & 1.51$\pm0.07$ & 0.27$\pm0.03$ & 2.2$^{+1.1}_{-0.8}$ & 1.17$\pm0.03$ & 25.1$\pm1.2$ & 9.85$\pm0.08$ & 87/96\\
        &0.43--0.54 & 3.0 & 8.4$\pm0.5$ & $1.61\pm0.06$ & 0.29$\pm0.03$ & 3.9$^{+1.6}_{-1.3}$ & 1.13$\pm0.02$ & 27.8$\pm1.3$ & 10.64$\pm0.08$ & 86/98\\
        &0.54--0.64 & 3.0 & 9.1$\pm0.7$ & 1.58$\pm0.07$ & 0.27$\pm0.03$ & 6.1$^{+2.9}_{-2.4}$ & 1.14$\pm0.02$ & 28.3$\pm1.5$ & 11.20$\pm0.09$ & 95/97\\
        &0.64--0.92 & 2.5 & 8.4$\pm0.5$ & $1.47\pm0.08$ & 0.23$\pm0.03$ & 4.0$^{+1.7}_{-1.4}$ & 1.17$\pm0.02$ & 27.2$\pm1.3$ & 11.58$\pm0.1$ & 94/97 \\
        &0.92--1.10 & 2.0 & 9.0$^{+0.3}_{-0.4}$ & 1.50$^{+0.08}_{-0.09}$ & 0.20$\pm0.02$ & 6.1$^{+1.6}_{-1.6}$ & 1.18$\pm0.01$ & 28.9$\pm1.0$ & 12.57$\pm0.1$ & 91/96\\
        &1.50--1.85 & 4.1 & 9.5$\pm0.4$ & 2.29$\pm0.06$ & 0.29$\pm0.02$ & 7.3$^{+1.8}_{-1.7}$ & 1.03$\pm0.02$ & 32.3$\pm1.1$ & 8.17$\pm0.08$ & 81/98 \\
        &1.85--1.97 & 3.0 & 8.3$^{+0.5}_{-0.4}$ & $2.50\pm0.01$ & 0.26$\pm0.03$ & 3.5$^{+1.5}_{-1.1}$ & 1.04$\pm0.02$ & 30.1$^{+1.4}_{-1.2}$ & 6.97$\pm0.07$ & 67/96 \\
        &1.97--2.10 & 3.1 & 8.2$\pm0.4$ & $2.70\pm0.1$ & 0.26$\pm0.03$ & 3.0$^{+1.3}_{-1.0}$ & 1.04$\pm0.02$ & 29.3$^{+1.4}_{-1.2}$ & 6.58$\pm0.07$ & 78/96\\
        &2.10--3.00 & 3.1 & 8.9$\pm0.4$ & 3.32$\pm0.1$ & 0.39$\pm0.06$& 6.3$^{+1.6}_{-1.5}$ & 1.07$\pm0.02$ & 29.7$\pm1.0$ & 7.82$\pm0.07$ & 80/97\\
        &3.00--4.00 & 0.9 & 7.3$\pm0.4$ & 2.1$^{fixed}$ & 0.020$\pm0.003$ & 1.8$^{+1.4}_{-0.9}$ & 1.21$\pm0.01$ & 24.1$\pm0.5$ & 6.63$\pm0.08$ & 89/94\\
        
        \hline
    \end{tabular}

\begin{flushleft}
Notes: \\
$\S$: The \texttt{XSPEC} model \texttt{tbabs*(diskbb+thcomp$\otimes$bbodyrad}) is applied to all the segments.\\
$\ast \ast$: LAXPC exposure time (ks).\\
$\ddag$: Value for the hydrogen column density ($10^{22}\mathrm{cm}^{-2}$).\\
$\P$: Photon index for \texttt{thcomp} model.\\
$\bullet$: Covering fraction parameter for \texttt{thcomp} model.\\
$\ast$: Defined as $(R_{\rm in}/D_{10})^2 \cos\theta$, where $R_{\rm in}$ is an ``apparent" inner disk radius in km, $D_{10}$ the distance to the source in units of 10\, kpc, and $\theta$ the angle of the disk ($\theta= 0$ is face-on) ($10^4$).  \\
$\star$: temperature of blackbody (keV). \\
$\dagger$: Defined as $D_{10}\times\sqrt{norm}$, where norm is the blackbody normalization value and $D_{10}$ is the distance to the source in units of 10 kpc (km). The source distance is taken to be 11\,kpc \citep{2000MNRAS.317....1F} \\ 
$\#$: Unabsorbed flux computed in the energy range 3.0--20.0\,keV ($10^{-9}$ $\mathrm{ergs~cm^{-2}s^{-1}}$) as estimated using cflux (cflux*(diskbb+thcomp$\otimes$bbodyrad)).\\
 
  \end{flushleft}
        
\end{table*}

\section{Discussion} \label{discussion}

In this work, we investigate the evolution of spectral and timing properties of the Z source GX 340+0 along different positions of the Z track using broadband X-ray observations with \textit{AstroSat}. The spectra of all $S_z$ segments of the Z track are consistently described by the same model \texttt{tbabs*(diskbb+thcomp$\otimes$bbodyrad)}, comprising thermal emission from the accretion disk and the neutron star (NS) surface, represented by a blackbody associated with the spreading boundary layer, together with a Comptonizing corona that up-scatters soft seed photons. While the disk contributes to the overall soft thermal emission, the seed photons for Comptonization are inferred to originate primarily from the neutron star surface rather than directly from the accretion disk. The fact that the same spectral model successfully describes the data across all $S_z$ segments suggests that the fundamental accretion geometry remains broadly similar throughout the Z track, while the observed spectral evolution is driven mainly by changes in parameters associated with the mass accretion rate and the structure of the spreading layer (SL)–corona system. This consistency allows us to interpret the spectral variations in a unified framework and to explore how the physical conditions in the inner accretion flow, the SL, and the surrounding Comptonizing region evolve as the source moves along the Z track. Note that our interpretation is primarily based on the SL framework of \citet{1999AstL...25..269I}, in which accreted material arriving from the inner disk spreads over the NS surface and increases its latitudinal extent with increasing accretion rate. The observed evolution of the blackbody normalization is interpreted within this framework as a consequence of changes in the extent of SL. However, the absolute normalization values inferred from our spectral fits correspond to emitting radii larger than the canonical NS radius, suggesting that the observed thermal emission may not arise exclusively from the stellar surface. In addition to the SL, a contribution from the boundary layer (BL), where the accreting matter decelerates from nearly Keplerian motion before settling onto the star, may increase the effective emitting area inferred from the blackbody component. In this picture, the variability of the thermal emission is governed primarily by the evolving SL, while BL provides an additional, approximately steady contribution to the observed blackbody normalization.

\subsection{Evolution along the Z track} \label{subsec:spectral_evolution}
\begin{figure*}
    \centering
\includegraphics[width=0.495\columnwidth]{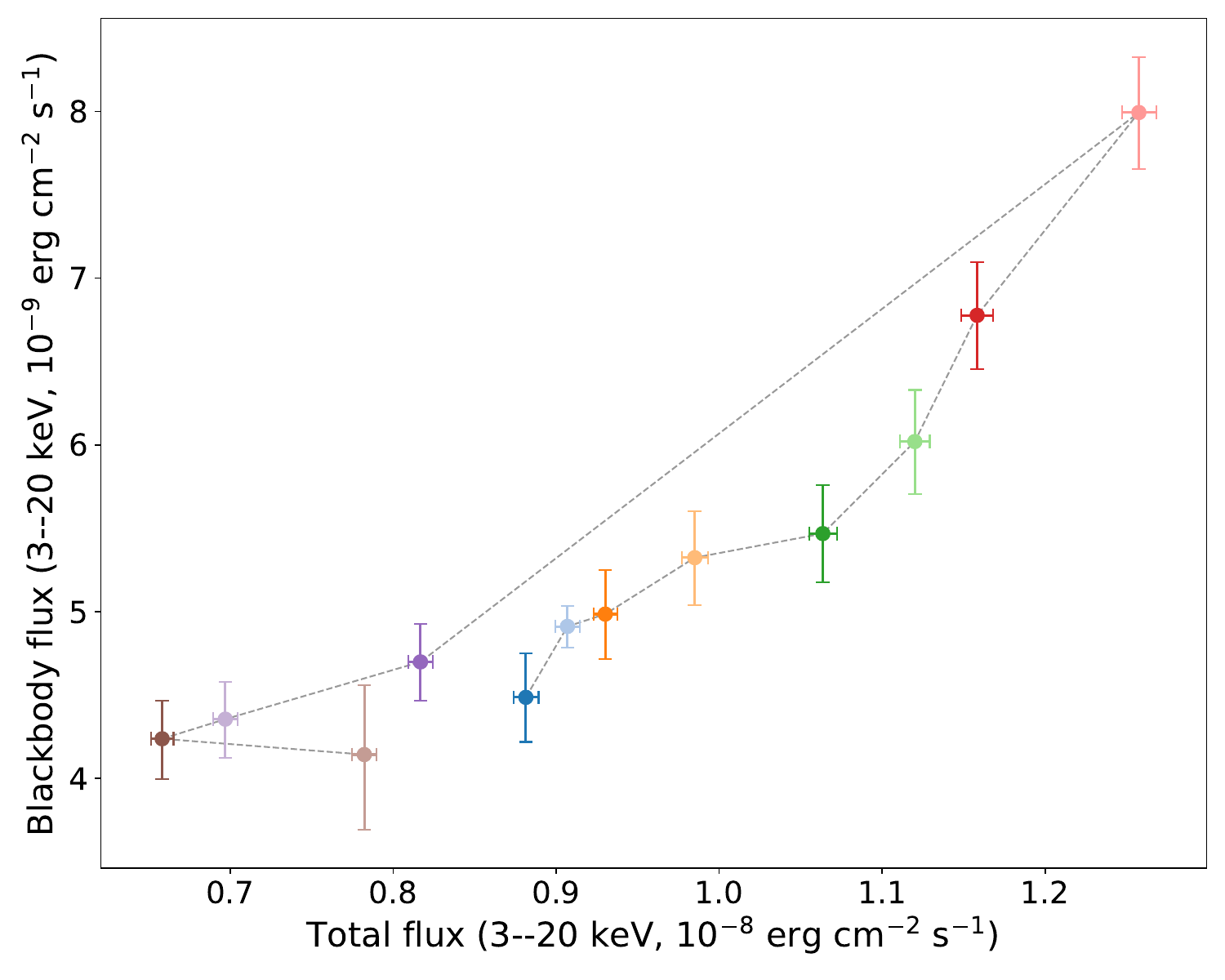}
\includegraphics[width=0.495\columnwidth]{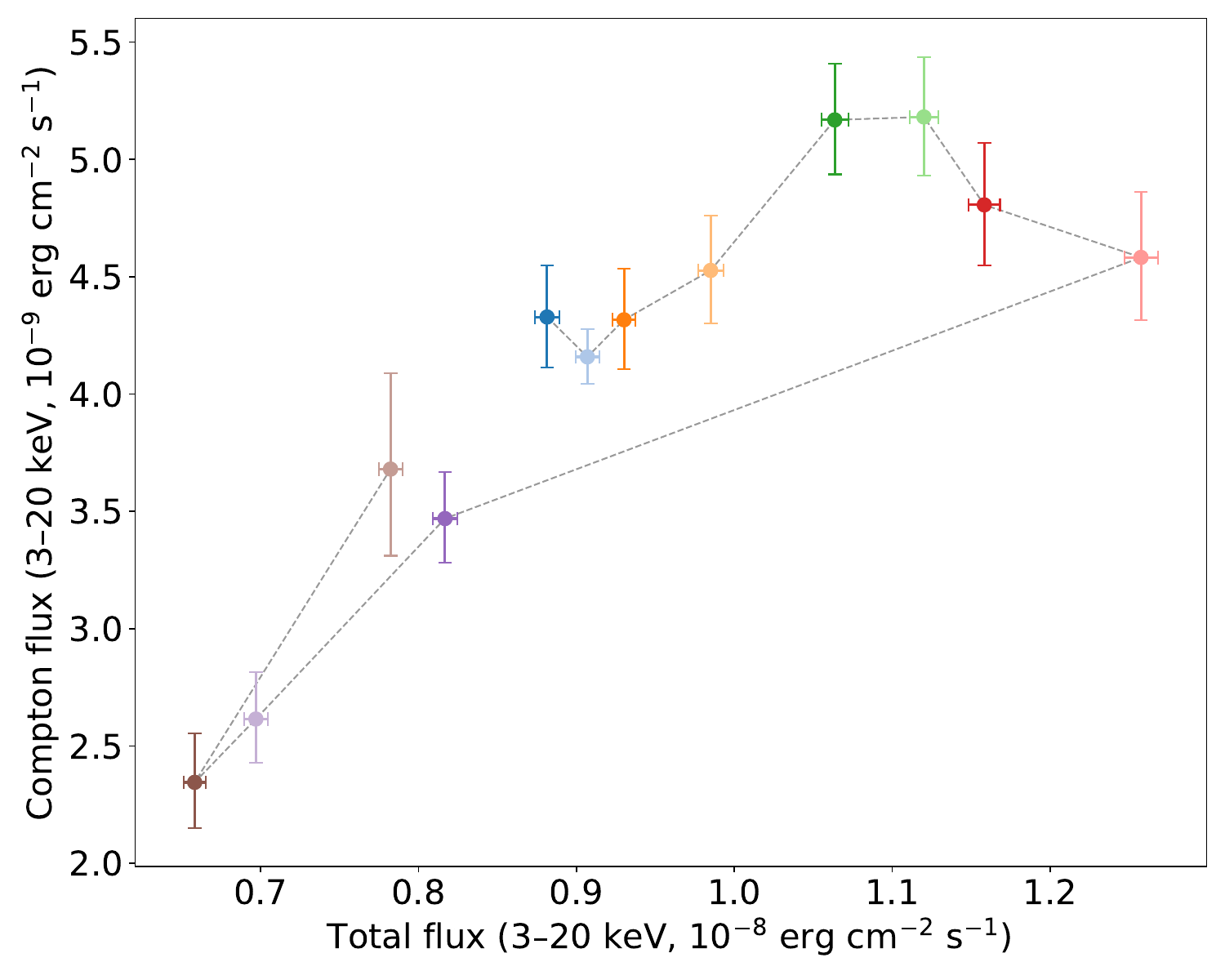}
\caption{The Figure shows the evolution of blackbody flux (left panel) and Comptonized flux (right panel) with total flux in 3--20\,keV energy range for the 2024 GX~340+0 \textit{AstroSat} data. The color scheme is kept the same as $S_z$ resolved HID, Figure~\ref{fig:lightcurve_hid}.}
    \label{fig:flux_evo}
\end{figure*}

Starting from $S_z = 0$, the accretion rate $\dot{M}$ is at the lower end of its variation along the horizontal branch (HB), with a corresponding unabsorbed flux of $\sim9\times10^{-9}$ erg cm$^{-2}$ s$^{-1}$ in the 3--20 keV band (see Figure~\ref{fig:flux_evo}). The accreting material arriving from the inner disk at nearly Keplerian velocity decelerates in BL before settling onto the NS surface. A fraction of this material subsequently spreads over the stellar surface, forming an equatorial belt as described in the SL framework of \citet{1999AstL...25..269I}. At low accretion rates, the limited latitudinal extent of the SL results in a relatively smaller effective emitting area and hence a lower blackbody normalization ($\sim$450), corresponding to an apparent emitting radius of $\sim$23\,km (Figure~\ref{fig:parameter_evo}; Table~\ref{tab:spec_pars}). However, the relatively large normalization values inferred from our spectral fits need not imply emission confined strictly to the stellar surface. Instead, while the observed evolution of the blackbody normalization is interpreted in terms of changes in the latitudinal extent of the SL, the absolute normalization may include contributions from both the SL and the BL. Such behaviour is broadly consistent with earlier studies showing that the extent of the emitting region is closely linked to the structure of the inner accretion flow and can exceed that expected from a purely surface-confined emitting layer \citep{2002A&A...390..139C}.
In this scenario, a geometrically thick, hot inner flow or coronal envelope may also develop above and around the inner disk–BL interface. Soft photons originating from the NS surface, primarily from the SL emission, are subsequently intercepted and Compton up-scattered by this surrounding coronal plasma, producing the hard Comptonized component and the significant covering fraction typically observed on the HB \citep{1999AstL...25..269I,2006A&A...460..233C}. In many earlier spectral interpretations of NS LMXBs, the seed photons for Comptonization have often been assumed to originate from the accretion disk and subsequently undergo Compton up-scattering in a hot corona \citep{2004MNRAS.348..955C,2009A&A...498..509F,2010A&A...512A..57M}. In the present interpretation, however, the seed photons are assumed to originate primarily from thermal emission associated with the NS surface. These two viewpoints may not be fundamentally different, since the BL and SL are ultimately fed by accretion disk material that reaches the stellar surface while still carrying nearly Keplerian angular momentum and releases its kinetic energy through viscous dissipation \citep{1999AstL...25..269I}. Thus, the observed blackbody emission can still be regarded as the reprocessed radiation of accreted disk material. The key observable distinction lies in the spectral shape. A standard accretion disk produces a multi-temperature spectrum characterized by a radial temperature gradient, whereas emission from the NS surface is expected to be closer to a single-temperature blackbody. In Z sources, these components are typically distinct, with the blackbody component generally hotter than the disk emission \citep{2006A&A...460..233C}. In our case, the spectral fits favor the latter interpretation, indicating that the Comptonizing region primarily interacts with photons emerging from the NS surface through the thermal emission of the SL.

As the source moves along the HB toward the hard apex, the inferred increase in both blackbody and Comptonized flux (and the corresponding increase in total flux observed in the HID) indicates a rising $\dot{M}$, with the total unabsorbed flux reaching $\sim1.3\times10^{-8}$ erg cm$^{-2}$ s$^{-1}$ (Figure~\ref{fig:flux_evo}). Within the SL framework, the accreted material spreads to higher latitudes on the NS surface as $\dot{M}$ increases, leading to an increase in the emitting area and hence the blackbody normalization. Previous works have shown that the half-height of the blackbody emitting belt ($h$) is comparable to the scale height $H$ of the inner disk \citep{2002A&A...390..139C,2006A&A...460..233C}. Thus, expansion of the SL can enhance irradiation of the inner disk and modify its vertical structure, potentially producing a geometrically thicker, radiation-supported configuration \citep{1999AstL...25..269I,1999MNRAS.303..139D}. Such changes may affect the effective low-energy absorption and alter the apparent disk normalization through changes in vertical structure and spectral hardening rather than simple radial truncation \citep{1995ApJ...445..780S}. Since the material associated with the inner accretion flow is expected to be highly ionized, the observed absorption may contain both interstellar neutral absorption and an intrinsic ionized contribution. To examine this possibility, we additionally tested an ionized absorber ({\tt zxipcf}) while fixing the neutral {\tt tbabs} column to the minimum value obtained from our fits. The resulting ionized column density shows a broadly similar qualitative evolution to that obtained with {\tt tbabs}, but the limited low energy coverage and its moderate spectral resolution leave the ionized absorber poorly constrained, resulting in large uncertainties. We therefore regard the observed $n_\mathrm{H}$ variation as being consistent with changes in the effective inner disk obscuration, but not as direct evidence for structural evolution of the disk.
As $\dot{M}$ increases further, the larger emitting area of SL and the enhanced energy dissipation in the boundary region where the disk flow settles onto the NS increase the supply of soft seed photons. Such an enhanced soft-photon field is expected to increase the efficiency of inverse Compton cooling of the surrounding hot plasma, potentially reducing the temperature and/or geometrical extent of the Comptonizing region \citep{1991ApJ...380L..51H,1997ApJ...476..620H,2007A&ARv..15....1D}. In our analysis, however, the electron temperature could not be reliably constrained across all $S_z$ segments because of its strong covariance with the photon index and covering fraction. We therefore fixed $\mathrm{k}T_{\rm e}=3.1$ keV, corresponding to the average value obtained from those spectra in which it was reliably constrained. Consequently, we interpret the observed decrease in the covering fraction toward the hard apex (Figure~\ref{fig:parameter_evo}) as being consistent with a reduction in the effective extent and/or scattering efficiency of the Comptonizing region, rather than as direct observational evidence for a decrease in the electron temperature.

As the source moves along the NB toward the soft apex, the decline in both blackbody and Comptonized flux implies a reduction in $\dot{M}$, consistent with the interpretations in which the accretion rate (or accretion rate per unit area) decreases toward the soft apex \citep{2006A&A...460..233C, balucinska2010A&A...512A...9B}, (Figure~\ref{fig:flux_evo}). The spreading layer contracts, and the already accreted material redistributes over a larger fraction of the NS surface (Figure~\ref{fig:parameter_evo}), producing a geometrically thinner emitting layer with lower temperature ($kT \sim 1.0$ keV) and larger blackbody normalization, while the gradually settling disk atmosphere leads to a reduction in $n_\mathrm{H}$ \citep{1999AstL...25..269I}. A similar decrease in the blackbody temperature toward softer states along the Z-track has been reported in earlier studies of Z-sources \citep{2002A&A...390..139C,2006A&A...460..233C}, and more recently in broadband spectral analyses of GX 340+0 and other systems \citep{2024MNRAS.528.6167C, 2025ApJ...994..169L}. 
The lower blackbody flux allows the corona to re-energize and expand geometrically, increasing the covering fraction but decreasing its optical depth, thereby softening the Comptonized spectrum (higher $\Gamma$). Finally, as the source enters the flaring branch (FB), renewed increase in $\dot{M}$ re-expands the SL, raises the local radiation pressure toward the Eddington limit, reinflates the inner disk, and enhances both blackbody and Comptonized emission, completing the cyclical restructuring of the disk–SL–corona system that maintains the source on the characteristic Z track \citep{1999AstL...25..269I, 2006A&A...460..233C, 2013MNRAS.434.2355R}.

Because the EFB segment has a short exposure ($\sim$900 s), the spectral parameters are not well constrained. To obtain reasonable estimates of the remaining parameters, we fix the photon index $\Gamma$ to the value reported by \citet{2024MNRAS.535.3383D}. The corresponding best-fit parameters are listed in Table~\ref{tab:spec_pars}, but are not included in the spectral evolution plots due to their limited statistical significance. For a detailed analysis of the EFB, we refer the readers to \citet{2024MNRAS.535.3383D}.

\subsection{Resolving the Hysteresis in Z-track}

Considering the discussion so far within the SL framework, we find that although the mass accretion rate $\dot{M}$ decreases as the source moves away from the hard apex along both the HB and NB, the subsequent spectral evolution differs significantly. Here we describe key physical differences between adjacent branches and their observables. Along the HB, the blackbody temperature remains relatively high ($\mathrm{k}T_{\mathrm{bb}}\sim1.2$ keV) and the Comptonized flux stays nearly constant, while the photon index $\Gamma$ remains flat.(Figures~\ref{fig:parameter_evo} and \ref{fig:flux_evo}). In contrast, along NB, the blackbody temperature decreases ($\mathrm{k}T_{\mathrm{bb}}\sim1.0$ keV), $\Gamma$ steepens, and the Comptonized flux declines significantly. Although the covering fraction increases along both branches (moving from $S_z$=1 to 0 for HB and 1 to 2 for NB), these trends suggest that the coupling between the SL and the Comptonizing region evolves differently along HB and NB. Within our interpretation, the relatively hotter thermal emission on HB provides a more energetic soft-photon field that sustains efficient Comptonization. In contrast, the lower blackbody temperature and higher $\Gamma$ observed along NB are accompanied by a significant reduction in the Comptonized flux, suggesting that the production of the hard Comptonized component becomes less efficient, possibly due to a softer seed photon field and a less effective Comptonizing region. Consequently, the spectrum remains relatively hard along HB but becomes progressively softer along NB, even at comparable luminosities. The evolution of the blackbody flux on both branches remains similar, Figure~\ref{fig:flux_evo} (left panel).

Similarly, as the source evolves away from the soft apex, the spectral evolution along FB and NB shows distinct behaviour despite a likely increase in $\dot{M}$ along both branches. Along FB, the increase in accretion rate leads to an expansion of SL to higher latitudes, resulting in a modest rise in blackbody temperature while the emitting area remains nearly constant. This is accompanied by an increase in $\Gamma$ and covering fraction, indicating a more Compton-cooled and geometrically extended corona. In contrast, along the NB (from $S_Z$=2 to 1), although $\dot{M}$ also increases, the SL spreads more gradually, leading to an increase in blackbody normalization at nearly constant temperature, while $\Gamma$ decreases, suggesting a relatively hotter and more efficient Comptonizing region. Although both $n_\mathrm{H}$ and disk normalization increase along both branches, the differing evolution of $\Gamma$ and covering fraction implies stronger SL–corona coupling toward FB compared to NB. Consequently, the spectrum becomes softer along FB, while remaining relatively harder along NB.

\subsection{Secular shift of the Z-track}

The secular shift observed between the 2020 and 2024 HIDs of GX~340+0 can be interpreted within the SL framework as a result of long-term variations in the average mass accretion rate. In both epochs, the source traces the characteristic Z-shaped track, and the same spectral model describes all $S_z$ segments, indicating that the overall accretion geometry comprising a Keplerian disk extending close to the NS surface, a SL producing thermal emission, and a surrounding Comptonizing corona, remains broadly unchanged. The secular displacement therefore reflects a modification of global accretion conditions rather than a change in the fundamental structure of the system.
The spectral evolution suggests that the 2020 observations correspond to a higher long-term accretion state than in 2024. In this picture, a higher $\dot{M}$ leads to SL extending to higher latitudes, increasing the emitting area and blackbody normalization, while lowering the effective temperature due to redistribution of the radiative flux. In contrast, during 2024 the SL appears more confined, producing a smaller emitting area and slightly higher temperature.
Changes in the vertical structure of the inner accretion flow accompany this shift. The larger $n_\mathrm{H}$ and disk normalization in 2020 suggest a geometrically thicker inner disk, likely due to stronger irradiation from the extended SL. The covering fraction is also higher, indicating a more extended corona intercepting a larger fraction of soft photons.
Despite these differences, the evolution of spectral parameters with $S_z$ remains similar in both epochs, implying that the same sequence of accretion states operates along the Z track. Long-term changes in $\dot{M}$ primarily modify the SL extent, disk thickness, and coronal geometry, shifting the overall luminosity scale without altering the track morphology.

The direction of the HID shift can be understood from the systematic offset in spectral parameters. The 2020 data show higher blackbody normalization, $n_\mathrm{H}$, disk normalization, and photon index $\Gamma$, indicating a more extended SL, thicker disk, and a more Compton-cooled corona. Although the increase in $\Gamma$ implies an intrinsically softer spectrum, enhanced absorption, reprocessing, and a larger covering fraction redistribute the emission within the observed band. Consequently, the hardness ratio can shift to higher values despite the softer intrinsic spectrum, explaining the upward displacement of the Z track in the HID.

\subsection{Oscillations in SL paradigm} \label{subsec:Timing}

The geometrically thick inner flow, discussed in section~\ref{subsec:spectral_evolution} has been suggested as a possible site for the generation of low frequency QPOs through global oscillation or precession modes \citep{1998ApJ...492L..59S, 2010MNRAS.405.2447I, 2020MNRAS.491.3245F}.
If such a hot inner flow exists in the vicinity of the disk and SL transition region, its characteristic dynamical timescale would depend on its radial extent and location relative to the NS. As the accretion rate increases along the horizontal branch (HB), the SL is expected to spread to higher latitudes on the NS surface \citep{1999AstL...25..269I}, while the inner accretion disk becomes increasingly radiation-supported and vertically extended. These structural changes can also influence the properties of the surrounding hot inner flow. In particular, the region where the disk interacts with the BL may shift inward as the radiation field becomes stronger, with radiation-driven drag effects facilitating enhanced inflow \citep{1989ApJ...346..844W, 1992ApJ...385..642W, 2013ApJ...772...94W}. Since the characteristic dynamical frequencies scale approximately as $\nu \propto r^{-3/2}$, even modest inward shifts of this interaction region can lead to significant changes in the observed variability frequencies. For example, an increase in HBO frequency from $\sim 20$~Hz to $\sim 40$~Hz for GX~340+0 (see Table 3 in \citet{2024arXiv240519324B}) corresponds to a reduction in the characteristic radius by a factor of $(40/20)^{-2/3} \approx 0.63$, implying an inward shift of $\sim 35\%$ in radius. This provides a natural explanation for the observed increase in HBO frequency along the HB. Additionally, since the HBO variability is believed to originate in the dynamically active inner flow or corona, changes in its geometry can significantly affect the observed amplitude and coherence of the oscillations. In our analysis, the photon index $\Gamma$ remains approximately constant along HB, indicating that the effective optical depth of the Comptonizing region does not vary significantly. However, we observe a decrease in the covering fraction as the source approaches the hard apex, suggesting a reduction in the geometrical extent or coupling of the corona with the soft photon source. In such a scenario, a smaller fraction of seed photons participate in the Comptonizing region responsible for the variability, leading to a dilution of the modulated emission and hence a decrease in the observed QPO rms. Thus, within the SL framework, the increase of HBO frequency along the HB can still be associated with an inward shift of the disk–BL interaction region, while the simultaneous decrease in rms reflects a reduced geometrical coupling and modulation efficiency of the corona, rather than a change in its optical depth. The fact that HBO frequencies occur at similar values for a given $S_z$ across observations further suggests that $S_z$ traces a characteristic radial scale of the inner flow, linking the QPO frequency to a preferred radius in the disk–BL interaction region.

The increase in hard lags toward the hard apex, as reported for GX 340+0 by \citet{2023ApJ...955..102B}, can be understood in the context of a complex accretion geometry rather than a simple change in coronal size or optical depth. In our analysis, the photon index $\Gamma$ remains approximately constant, and with the electron temperature fixed ($k\mathrm{T}_e = 3.1$ keV), this suggests that the effective optical depth of the Comptonizing region does not vary significantly. This indicates that the observed lag evolution is unlikely to be driven solely by changes in scattering depth. Instead, as the accretion rate increases along the HB, the SL expands and enhances its coupling with both the inner disk and the Comptonizing region. As the disk–BL interaction region moves closer to the neutron star surface, the effective emission region becomes more deeply embedded within this coupled structure, such that photons may undergo additional diffusion and multiple scattering interactions before escaping, increasing their effective path length, similar to the Compton diffusion scenarios proposed for extended corona \citep{2000astro.ph..1242K} and consistent with the interpretation of energy-dependent lags arising from Comptonization processes \citep{1999ApJ...517..355N}. In such a configuration, variability may propagate through radially and vertically stratified regions of the flow, introducing delays as fluctuations travel inward \citep{1997MNRAS.292..679L,2001MNRAS.327..799K,2006MNRAS.367..801A}. In addition, radiative feedback between the SL and the corona can further increase photon travel times, as soft photons emitted from the NS surface are Comptonized, reprocessed, and re-injected into the hot flow \citep{2001ApJ...549L.229L}. The optically thick nature of the SL itself may also introduce diffusion delays prior to photon escape \citep{1999AstL...25..269I}. These combined effects may lead to increase in hard lags without requiring a significant change in the average coronal optical depth. The simultaneous evolution of lag amplitude and spectral properties therefore suggests that the inner accretion flow becomes more strongly coupled and stratified toward the hard apex, providing a consistent explanation of the observed timing behavior within the SL--corona framework.

\subsection{Origin of polarization in HB and NB in SL paradigm}

The recent IXPE polarization results of GX~340+0
\citep{2024arXiv240519324B,2024arXiv241100350B,
2024A&A...691A.253L,2025A&A...702A.101L} provide important constraints on the geometry of the inner accretion flow. More generally, a systematic model-independent analysis of IXPE observations of Z-sources has shown that the polarization degree is typically higher on the HB and decreases toward the softer branches, while the polarization angle varies only modestly along the Z track
\citep{2025A&A...699A.230G}. The observed decrease in polarization degree from $\sim4\%$ on the HB to $\sim1.3\%$ on the NB in GX~340+0 therefore suggests that the physical conditions within the inner accretion flow evolve significantly as the source moves along the Z track.

Our spectral decomposition provides a semi-quantitative connection between the X-ray spectral components and the IXPE observations. Since X-ray polarization depends on the number of detected photons, we estimate the photon flux contributed by each spectral component in the IXPE 2--8 keV band using the {\tt cpflux} convolution model. Within our partial covering geometry, only a fraction of the seed blackbody photons escapes directly, while the remaining fraction is intercepted by the Comptonizing medium. The direct blackbody and Comptonized photon fluxes are therefore reconstructed using the best-fit covering fraction together with the photon fluxes returned by {\tt cpflux}. The resulting average photon flux fractions are approximately $69\%$, $17\%$, and $14\%$ for the direct SL emission, Comptonized emission, and disk emission, respectively, along the HB, while along the NB they become approximately $58\%$, $23\%$, and $19\%$. Thus, the relative photon contributions of the individual spectral components evolve only moderately despite the observed factor of $\sim3$ decrease in the measured polarization degree.

The observed polarization is determined by the vector addition of the Stokes parameters of the individual spectral components and therefore depends on their photon fluxes, intrinsic polarization degrees, and polarization angles. Recent spectropolarimetric modeling of GX~340+0 indicates that the
Comptonization/reflection and disk components differ in polarization angle by
approximately $30^\circ$--$40^\circ$ \citep{2024arXiv240519324B,2024arXiv241100350B, 2025A&A...699A.230G}, implying that the system is not perfectly azimuthally symmetric and that partial cancellation of the polarized Stokes vectors may occur in the energy range where these components overlap. Consequently, changes in the relative photon fluxes alone cannot account for the observed evolution of the net polarization.

A more significant evolution is instead seen in the properties of the Comptonizing medium itself. While the electron temperature remains fixed ($\mathrm{k}T_{\rm e}=3.1$ keV), the inferred Compton optical depth decreases systematically from $\tau\sim17$--18 on the HB to $\tau\sim8$ on NB and continues to decline toward the softer branches\citep{2023ApJ...955..102B}, see figure~\ref{fig:appendix_fig2}. In contrast, the covering fraction shows only modest variations, decreasing slightly along the HB before increasing again across the NB (figure~\ref{fig:parameter_evo}). This suggests that although a comparable fraction of seed photons may continue to enter the Comptonizing region, the photons experience substantially fewer scatterings before escaping as the source evolves toward softer states. Since multiple anisotropic Compton scattering events together with disk reflection are expected to generate most of the polarized emission \citep{2022MNRAS.514.2561G,2025A&A...699A.230G,2025A&A...702A.272F}, the reduction in optical depth naturally decreases the intrinsic polarizing efficiency of the hot inner flow. We therefore conclude that the observed decrease in polarization degree is primarily driven by changes in the scattering properties of the Comptonizing medium rather than by the relatively modest redistribution of photon flux among the thermal, Comptonized, and disk components.
The SL itself is expected to produce only weakly polarized radiation ($\lesssim1.5\%$) because of its large optical depth and nearly isotropic emission
\citep{2024A&A...684A..62F,2025A&A...696A.181B}. Consequently, although the SL dominates the observed photon flux in the IXPE band, it is unlikely to dominate the polarized signal. Instead, the observed polarization is most naturally explained by the hot Comptonizing flow and its interaction with the accretion disk. Within this picture, the decline in polarization from the HB to the NB primarily reflects a reduction in the intrinsic polarizing efficiency of the scattering medium, rather than a simple redistribution of photons among the thermal, Comptonized, and disk components.

More generally, the observed polarization is likely determined by the combined contributions of weakly polarized thermal emission, Comptonized emission, disk reflection, and potentially scattering in optically thin material. Recent XRISM observations of GX~340+0 have revealed evidence for an
accretion-disk wind, suggesting that scattering in outflowing material may also contribute to the observed polarization signal \citep{2025ApJ...991..224C}. Within this framework, the increase in HBO
frequency along the HB reflects an inward shift of the dynamically active inner flow, while the subsequent evolution toward the NB corresponds to a reconfiguration of the corona--SL system that simultaneously reduces both the intrinsic polarization of the scattering medium and the
rapid X-ray variability.

\begin{figure*}
    \centering
\includegraphics[width=0.49\columnwidth]{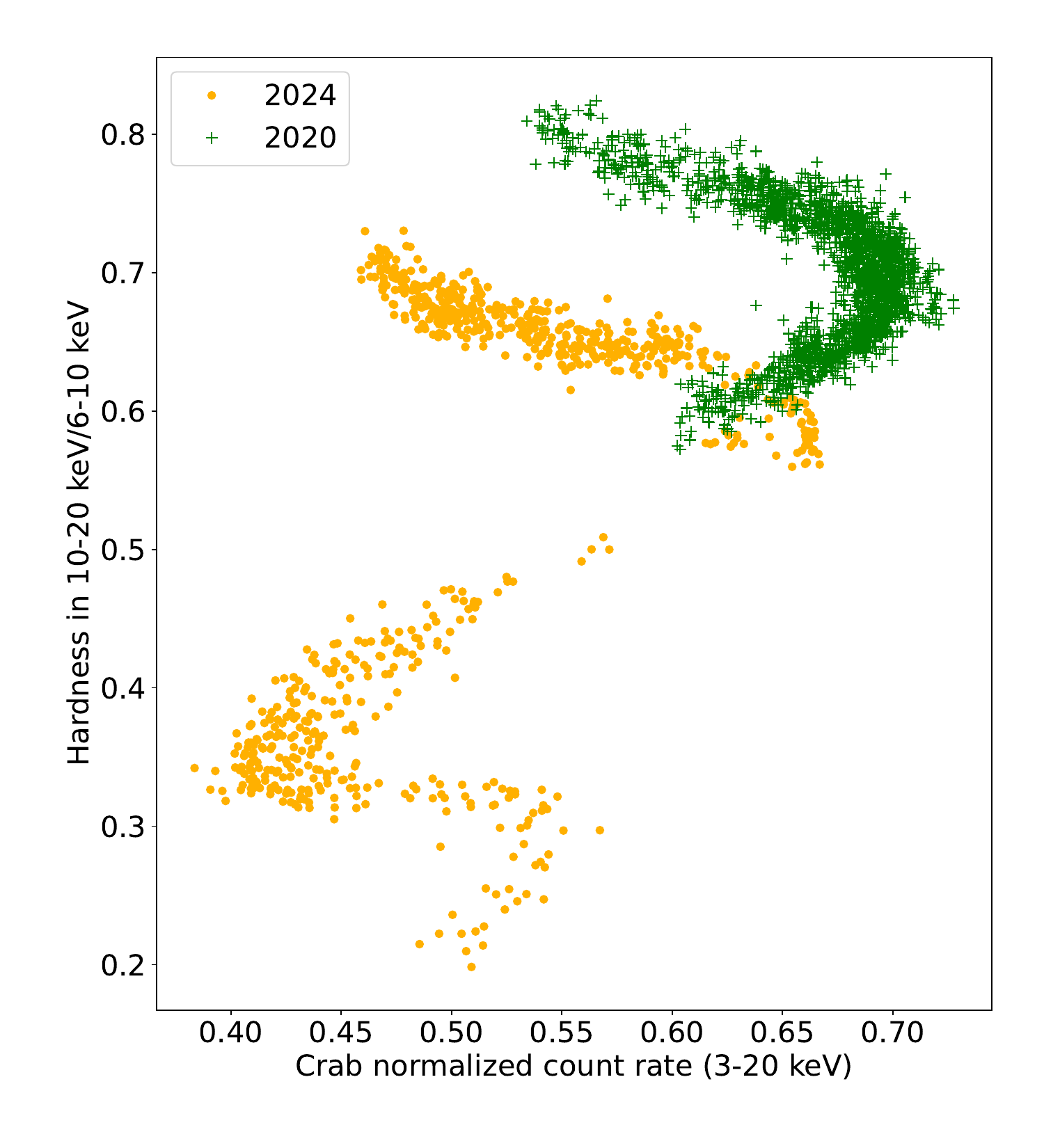}
\includegraphics[width=0.49\columnwidth]{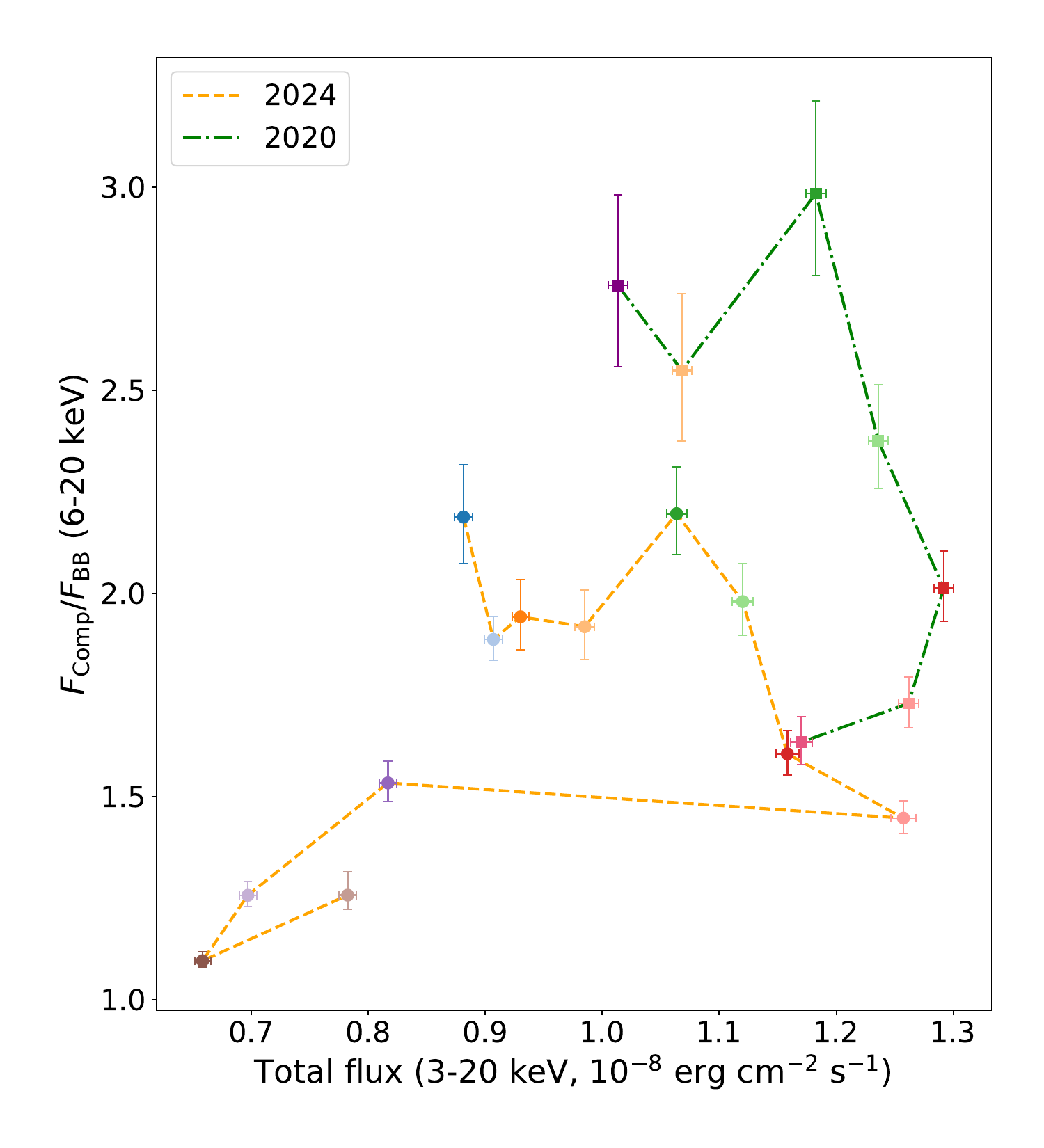}
\caption{Left panel: the hardness intensity diagram (HID) of source GX 340+0 observed in 2020 and 2024 by \textit{AstroSat}. The plot shows a secular shift in the HID overtime. Right panel: the ratio of the Comptonized flux to the blackbody flux in 6–20 keV as a function of the total flux in 3–20 keV. This panel reproduces the HB, the hard apex, and the NB of the observed Z track qualitatively.}
    \label{fig:secular_shift}
\end{figure*} 

\section{Summary}\label{sec:summary}

\begin{itemize}

\item We find that the broadband X-ray spectrum of GX~340+0 across the Z track is well described by a model comprising thermal emission from both the accretion disk (\texttt{diskbb}) and the neutron star surface (\texttt{bbodyrad}, associated with a spreading layer), together with a Comptonizing corona (\texttt{thcomp}). In this framework, the seed photons for Comptonization are assumed to originate primarily from the neutron star surface.

\item The Z track is parameterized using $S_z$ values, and the same spectral model is applied to all segments, enabling a uniform investigation of the evolution of spectral parameters and physical conditions as the source moves along the track.

\item The observed evolution of spectral parameters suggests that the coupled interaction between the spreading layer, corona, and inner accretion disk governs the motion of the source along the Z track. Variations in the spreading layer extent (traced by blackbody normalization), neutron star surface temperature, and coronal properties collectively regulate the spectral state.

\item We observe a significant decrease in blackbody temperature from HB to NB, which likely contributes to the harder spectra on the HB and softer spectra on the NB, even though the accretion rate decreases along both branches away from the hard apex. This highlights the role of the neutron star surface temperature in controlling the efficiency of Comptonization.

\item The variation in absorption column density suggests structural changes in the inner accretion disk, possibly linked to radiation driven puffing up of the disk at higher accretion rates, which may also influence the observed spectral and polarization properties.

\item A secular shift is observed between the 2020 and 2024 Z tracks, which may be attributed to long-term changes in the average accretion rate and corresponding adjustments in the geometry of the inner accretion flow, including the spreading layer, corona, and disk structure.

\end{itemize}

\begin{acknowledgements}
 This work makes use of data from the \textit{AstroSat} mission of the Indian Space Research Organisation (ISRO), archived at Indian Space Science Data Centre (ISSDC). The article has used data from the SXT and the LAXPC developed at TIFR, Mumbai, and the \textit{AstroSat} POCs at TIFR are thanked for verifying and releasing the data via the ISSDC data archive and providing the necessary software tools. This research has also made use of data and/or software provided by the High Energy Astrophysics Science Archive Research Center (HEASARC), which is a service of the Astrophysics Science Division at NASA/GSFC and the High Energy Astrophysics Division of the Smithsonian Astrophysical Observatory. 
\end{acknowledgements}

\begin{appendix}
\section{Assessment of the Disk Temperature Across the Z Track}
\renewcommand{\thesection}{\Alph{section}}
\setcounter{figure}{0}
\renewcommand{\thefigure}{A\arabic{figure}}
\label{appendix}

\begin{figure}[H]
    \centering
\includegraphics[width=0.55\columnwidth]{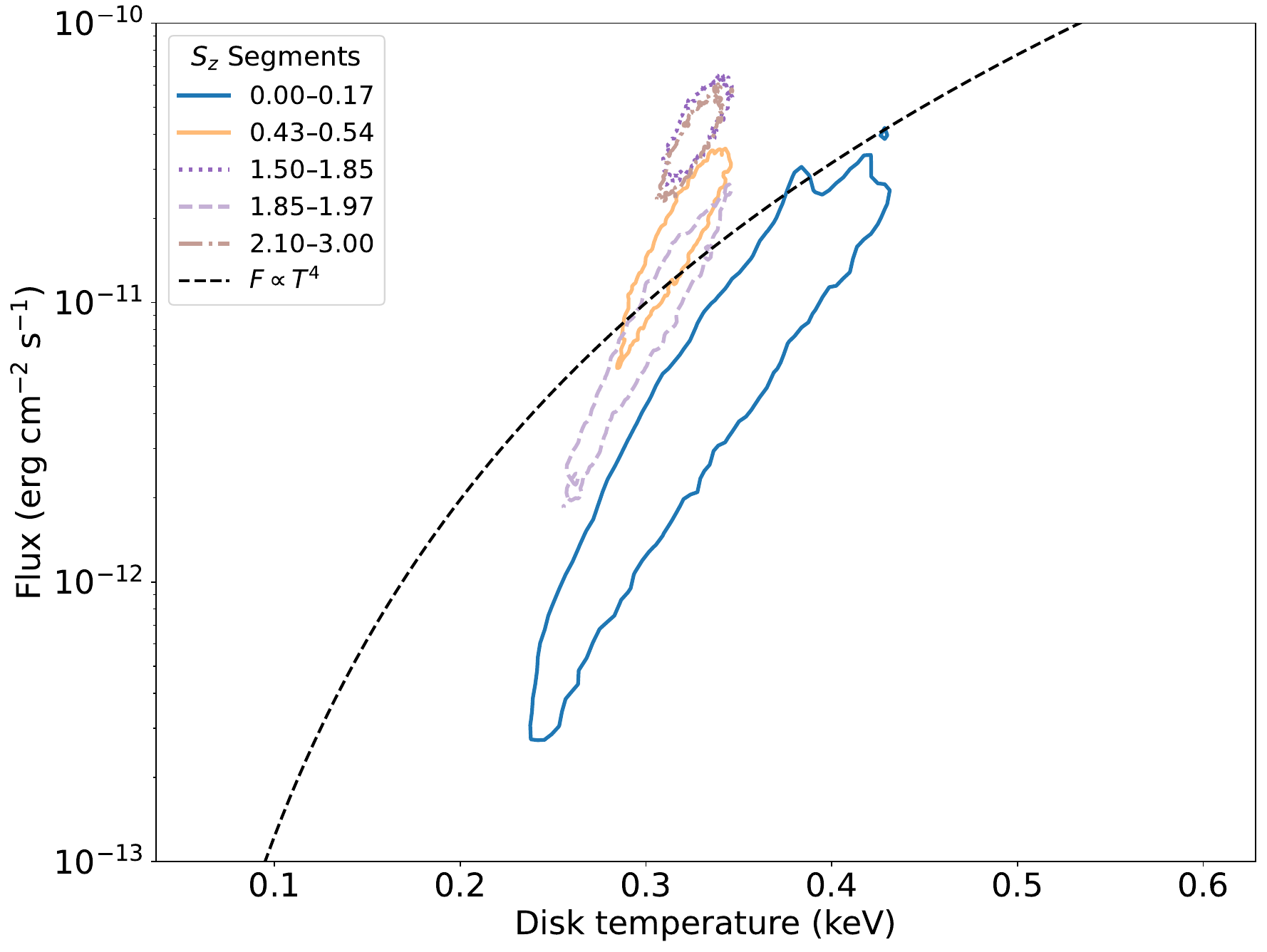}
\caption{Unabsorbed disk flux as a function of the best-fit disk temperature obtained with $\mathrm{k}T_{\rm in}$ allowed to vary. The color scheme is kept the same as $S_z$ resolved HID, Figure~\ref{fig:lightcurve_hid}. The dashed curve (black) indicates the expected $F_{\rm disk}\propto T_{\rm in}^{4}$ relation. While the disk flux evolves along the Z track, the disk temperature remains confined to a narrow range, indicating no significant systematic evolution of the characteristic disk temperature.}

    \label{fig:appendix_fig}
\end{figure}

To assess whether the disk temperature varies systematically along the Z track, we examined the evolution of the unabsorbed disk flux as a function of the best-fit disk temperature ($\mathrm{k}T_{\rm in}$) obtained from spectral fits with $\mathrm{k}T_{\rm in}$ left free. The resulting flux--temperature relation is shown in Figure~\ref{fig:appendix_fig}. Although the disk flux changes significantly across different $S_z$ segments, the inferred disk temperature remains confined to a narrow range of $\sim0.2$--0.4 keV. This behaviour indicates that the characteristic thermal state of the accretion disk remains nearly unchanged along the Z track. Since the peak of such a cool multi-colour disk lies largely below the observed energy range and is strongly affected by interstellar absorption, the spectra primarily sample only the high energy tail of the disk emission, making $\mathrm{k}T_{\rm in}$ only weakly constrained and strongly covariant with the disk normalization. Together with the simultaneous fitting of representative HB, NB, and FB spectra, which yielded a common best-fit temperature of $\mathrm{k}T_{\rm in}\simeq0.33$ keV, this motivated our choice to fix $\mathrm{k}T_{\rm in}=0.33$ keV in the final spectral analysis.\\

\section{Optical depth evolution along the Z curve}

\begin{figure}[H]
    \centering
\includegraphics[width=0.55\columnwidth]{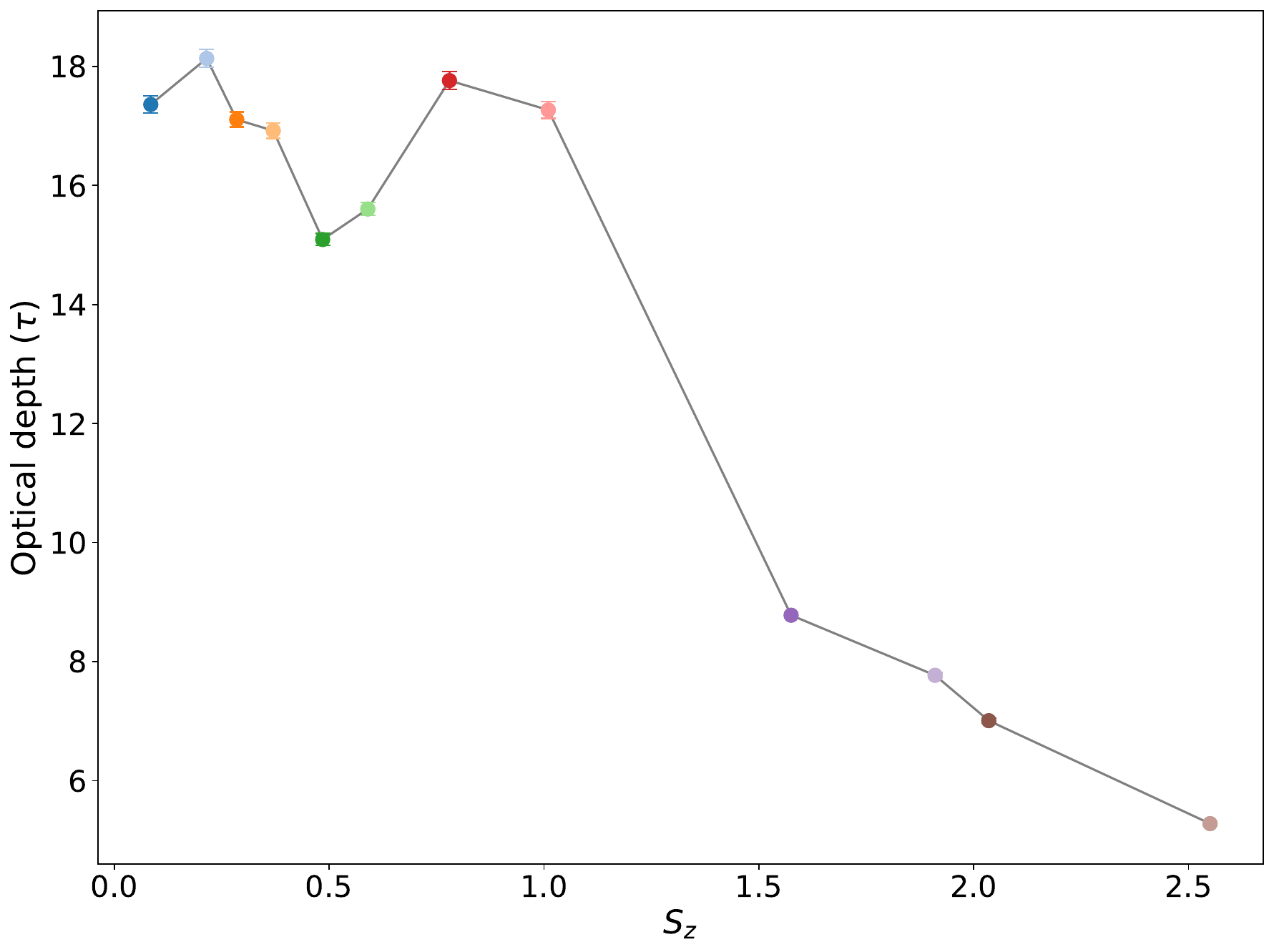}
\caption{Evolution of the Compton optical depth ($\tau$) of the Comptonizing region along the Z track, obtained from the {\tt thcomp} fits with $kT_{\rm e}=3.1$ keV fixed. The color scheme is kept the same as $S_z$ resolved HID, Figure~\ref{fig:lightcurve_hid}. The optical depth remains high, $\tau\sim15$--18, over the HB and decreases substantially toward the NB and FB, reaching $\tau\sim5$ on the FB. Note: in some data points the error bars are smaller than the marker size.}
    \label{fig:appendix_fig2}
\end{figure}

The optical depth of the Comptonizing region is estimated using the {\tt thcomp} model by fitting the spectra with the electron temperature fixed at $kT_{\rm e}=3.1$ keV. The resulting evolution of the Compton optical depth, $\tau$, along the Z track is shown in Figure~\ref{fig:appendix_fig2}. While $\tau$ shows modest variations within the HB, it decreases substantially as the source moves toward the NB and FB, from $\sim17$--18 on the HB to $\sim8$ near the HB--NB transition and $\sim5$ on the FB. This indicates a progressively lower scattering depth of the Comptonizing region toward the softer branches.

\end{appendix}

\bibliography{ref}
\bibliographystyle{aasjournal}

\end{document}